\documentclass[twocolumn,english,amsart,showpacs,preprintnumbers,amsmath,amssymb,floatfix]{revtex4-1}
\usepackage{tikz,xcolor}
\usepackage[colorlinks = true,
linkcolor = blue,
urlcolor  = blue,
citecolor = blue,
anchorcolor = blue]{hyperref}
\definecolor{lime}{HTML}{A6CE39}
\DeclareRobustCommand{\orcidicon}{%
	\begin{tikzpicture}
	\draw[lime, fill=lime] (0,0)
	circle [radius=0.16]
	node[white] {{\fontfamily{qag}\selectfont \tiny ID}};
	\draw[white, fill=white] (-0.0625,0.095)
	circle [radius=0.007];
	\end{tikzpicture}
	\hspace{-2mm}
}
\foreach \x in {A, ..., Z}{%
\expandafter\xdef\csname orcid\x\endcsname{\noexpand\href{https://orcid.org/\csname orcidauthor\x\endcsname}{\noexpand\orcidicon}}
}

\usepackage[T1]{fontenc}
\usepackage[latin9]{inputenc}
\usepackage{color}
\usepackage{array}
\usepackage{amstext}
\usepackage{graphicx}
\usepackage{esint}
\usepackage{rotating}
\usepackage{appendix}
\usepackage{float}

\usepackage[font=small,labelfont=bf]{caption}
\usepackage{xcolor}
\usepackage{ulem}

\makeatletter

\providecommand{\tabularnewline}{\\}

\@ifundefined{textcolor}{}
{%
 \definecolor{BLACK}{gray}{0}
 \definecolor{WHITE}{gray}{1}
 \definecolor{RED}{rgb}{1,0,0}
 \definecolor{GREEN}{rgb}{0,1,0}
 \definecolor{BLUE}{rgb}{0,0,1}
 \definecolor{CYAN}{cmyk}{1,0,0,0}
 \definecolor{MAGENTA}{cmyk}{0,1,0,0}
 \definecolor{YELLOW}{cmyk}{0,0,1,0}
 }

\@ifundefined{definecolor}
 {\usepackage{color}}{}
\@ifundefined{definecolor}
 {\usepackage{color}}{}
\makeatother
\usepackage{babel}
\begin{document}


\title{Nucleon spin structure functions, considering target mass correction and higher twist effects  at the  NNLO accuracy and  their transverse momentum dependence}

\author{ Abolfazl Mirjalili$^{1}$\orcidA{}}
\email{A.Mirjalili@yazd.ac.ir (Corresponding author)}

\author {Shahin Atashbar Tehrani$^{2}$\orcidB{}}
\email{Atashbar@ipm.ir}

\affiliation {
	$^{(1)}$Physics Department, Yazd University, P.O.Box 89195-741, Yazd, Iran       \\
	$^{(2)}$School of Particles and Accelerators, Institute for Research in Fundamental Sciences (IPM), P.O.Box
	19395-5531, Tehran, Iran}
\date{\today}
\begin{abstract}\label{abstract}
Using recent and updated world data on polarized structure functions $g_1$ and $g_2$ we perform  QCD analysis at the next-next-to-leading-order (NNLO) accuracy. We include also target mass correction and higher twist effect to get more precise results in our fitting procedure. To confirm the validity  of our fitting results several sum rules are examined and we do a comparison for them with  results from  other models. In our analysis we employ  Jacobi polynomials approach to obtain analytical solutions  of the DGLAP evolution equations for parton distribution functions (PDFs).
Using the extracted PDFs from our data analysis as input  we also
compute the $x$- and $\mathbf{p}_{T}$-dependence of some  transverse momentum dependence (TMD) PDFs in polarized case, based on covariant parton model.  These functions are
naively even time-reversal (T-even) at twist-2 approximation. The results for TMDs  are indicating proper and acceptable behaviour with respect to what are presented in other literatures.
\end{abstract}

\maketitle
\tableofcontents{}
\section{Introduction}\label{sec:intro}
The determination of the nucleon's spin into its quark and gluon components is still an important challenge in particle physics.
The deep-inelastic scattering (DIS) experiments performed at DESY, SLAC, CERN, and JLAB have refined our understanding of the spin distributions and revealed the spin-dependent structure functions of the nucleon. The polarized structure functions $g_1(x,Q^2)$ and $g_2(x,Q^2)$ are measured in deep-inelastic scattering of a longitudinally polarized lepton on polarized nuclear targets. {{} We do the required analysis on the polarized structure function  to extract the desired parton densities at the initial $Q_0$.

In exact consideration of inclusive processes it is required to take into account the distributions in which the role of transverse momentum is embedded. These distributions are known as transverse momentum dependent (TMD) distributions. TMDs are the generalization of PDFs which provide us an extensive knowledge to investigate the hadron structure function. {{}In a native  parton model  in which the effect of transverse momentum  of a quark is not outstanding, there is a proper computational frame  which is called infinite momentum frame (IMF) \cite{imf1,imf2}. In this frame the target (nucleon) is moving fast, comparable to speed of light and because of  Lorentz contraction the nucleon seems like a flat disc. In this case one can imagine a transverse space position of quark inside the disk with respect to the moving direction of target. This space coordinate is called  impact parameter and denoted usually by $b_T$. Corresponding to the impact parameter in coordinate space we can attribute to a quark inside the target a transverse momentum, $k_T$, that is perpendicular to moving direction of nucleon. This momentum  component is ignorable against the quark longitudinal momentum. This  model then gives oversimplified relations between structure and distribution functions. In an another model which is called {covariant parton model (CPM)~\cite{Landshoff:1971xb}} a more exact but much more complex relations between  structure and distribution functions are given. The original assumptions of this model is based on covariance of relations together with a spherically symmetric quark momenta distribution in the nucleon rest frame  where one photon exchange is used  in a charged lepton-quark interaction. The output of this model is  such us the quark transverse momentum is as important as longitudinal one  and the transverse momentum dependence  of parton densities are obtained analytically \cite{cpm1}.}

The extended PDF is then describing the parton distribution with respect to both $x$ and $k_T$ variables. On the other words quarks can have  transverse momentum with respect to the motion of parent hadron. The  transverse momentum of parton at initial state and inside the parent hadron  is called the intrinsic transverse momentum, denoted by $k_T$. In the final state  the transverse momentum of parton with respect to the momentum  of produced hadron is denoted by $p_T$. TMDs have outstanding effect on the momentum feature  of produced hadron. They also have  crucial role  to describe the spin asymmetry in produced hadron \cite{sa} by analysing the semi inclusive DIS (SIDIS) processes \cite{sa1,ours}. To achieve the  three dimensional (3D) picture of nucleon, some processes like SIDIS are required in which one can measure the effect of  transverse momentum of partons in created hadron. It is therefore required to consider the spin dependence of PDFs. {Early applications to polarized structure functions were made by~\cite{Jackson:1989ph,Roberts:1996ub,Blumlein:1996tp}}

The PDFs in polarized case are two types. The first one is related to the longitudinal polarized quark inside longitudinal polarized nucleon, denoted by $g_1(x)$ that is  called helicity function. The second one is related to transverse polarized quark inside the transverse polarized nucleon, denoted by $h_1(x)$ and is called the transversity function. The type of polarization is determined with respect to moving direction of nucleon. If the parton transverse momentum  as an extra degree of freedom  is also considered then total number of PDFs, involving polarized cases, are arising to eight ones \cite{sa2}.
In this article the polarized TMDs which are even time reversal functions, based on covariant parton model, are investigated.
}

The organization of this paper are as following. In Sec.\ref{structure-function}  an overview on theoretical aspects of  polarized structure function is done. In Sec.\ref{Jacobi-polynomials} the theoretical framework of  Jacobi polynomials approach is reviewed. Sec.\ref{target-mass-correction} is devoted to discuss the target mass correction for $g_1$ and $g_2$ structure functions. Additionally in Sec.\ref{Higher-twist-effect} higher twist effect is demonstrated for polarized structure functions. In Sec.\ref{QCD analysis and fitting procedure} which includes also some subsections we illustrate our QCD data analysis which we call it as MA22 analysis. To get more validation of our MA22 results, we examine in Sec.\ref{Sum-Rules} several sum rules. In Sec.\ref{helicity} our prediction for polarized PDFs and structure functions are presented. Using  the results of our MA22 analysis, some polarized TMDs can be calculated. We do it in Sec.\ref{TMD}. In the last part that is Sec.\ref{Summary} our conclusions is given.

\section{Leading twist spin dependence of structure function \label{structure-function}}

{{} To achieve the main goal of this article to calculate the polarized TMDs we first need  to analysis DIS structure function in polarized case. For this purpose linear combination of polarized parton densities and coefficient functions can be used to express the leading twist spin-dependent proton and neutron structure functions, $g_1^{\rm p} (x, Q^2)$ at the next-next-to-leading-order (NNLO) accuracy as it follows ~\cite{Shahri:2016uzl,Goto:1999by,Khanpour:2017cha}:}

\begin{eqnarray}\label{eq:g1pxspace}
	&&g_1^{\rm p} (x, Q^2) = \frac{1}{2} \sum_q e^2_q  \Delta q_{v}(x, Q^2)\otimes\nonumber\\ &&\left(1+\frac{\alpha_s(Q^2)}{2 \pi}\Delta C^{(1)}_q+\left(\frac{\alpha_s(Q^2)}{2 \pi}\right)^2\Delta C^{(2)}_{ns}\right)\nonumber\\
	&&+e^2_q (\Delta q_s+\Delta \bar{q_s})(x, Q^2)\otimes \nonumber \\
	&&\left(1+\frac{\alpha_s(Q^2)}{2 \pi}\Delta C^{(1)}_q+\left(\frac{\alpha_s(Q^2)}{2 \pi}\right)^2\Delta C^{(2)}_{s}\right)\nonumber\\
	&&+\frac{2}{9}\left(\frac{\alpha_s(Q^2)}{2 \pi}\Delta C^{(1)}_g+\left(\frac{\alpha_s(Q^2)}{2 \pi}\right)^2\Delta C^{(2)}_g\right)\otimes\Delta g(x, Q^2)\nonumber\\
\end{eqnarray}

Here $\Delta q_{v}$, $\Delta q_s$ and $\Delta g$  are the polarized valance, sea and
gluon densities, respectively. The pQCD evolution kernel for PPDFs is now available at the NNLO  {accuracy in Ref.~\cite{Moch:2014sna,Blumlein:2021ryt,Blumlein:2021enk}}.
The $\Delta C^{(1)}_q$ and $\Delta C^{(1)}_g$ {{} in Eq.(\ref{eq:g1pxspace}) are denoting} to the NLO spin-dependent quark and gluon hard scattering coefficients, calculable in  pQCD~\cite{Lampe:1998eu}.
We now apply the hard scattering coefficients, extracted at NNLO approximation.
At this order the Wilson coefficients are different for quarks and antiquarks {{}. They are presented in Eq.(\ref{eq:g1pxspace}) by $\Delta C^{(2)}_{ns}$ and $\Delta C^{(2)}_{s}$ respectively and their analytical relations  have been  reported in \cite{Zijlstra:1993sh}.
The symbol $\otimes$ in Eq.(\ref{eq:g1pxspace}) is representing typical convolution in $x$-space.

The neutron structure function, $g_1^{\rm n} (x, Q^2)$, can be obtained from the proton one by considering isospin symmetry. Hence the deuteron structure function at leading twist would be available, utilizing the $g_1^{\rm p}$ and $g_1^{\rm n}$  structure functions such as:}

\begin{eqnarray}\label{eq:g1dxspace}
	g_{1}^{\rm \tau 2(d)}(x,Q^2) = \frac{1}{2}\{g_1^{\rm p}(x,Q^2) + g_1^{\rm n}(x,Q^2)\} \times (1 - 1.5 w_D) \,,\nonumber  \\
\end{eqnarray}
where $w_D=0.05\pm0.01$ is the probability to find the deuteron in a $D-$state~\cite{Lacombe:1981eg,Buck:1979ff,Zuilhof:1980ae}. {{}
Using the {Wandzura and Wilczek (WW) relation~\cite{Wandzura:1977qf}} the leading twist polarized structure function of $g_{2} ^{\tau_2}(x, Q^2)$  can be fully determined via $g_1^{\tau_2}(x, Q^2)$ structure function:}
\begin{eqnarray}\label{eq:WW}
	g_{2} ^{\tau_2}(x, Q^2) & = & g_{2} ^{WW}(x, Q^2) =  \nonumber  \\
	&& - g_1^{\tau_2}(x, Q^2) + \int_x^1\frac{dy}{y} g_1^{\tau_2}(y, Q^2)\,. 
\end{eqnarray}
{{} This relation that is in the leading twist approximation can also be used when target mass correction (TMC) is included ~\cite{Wandzura:1977qf}.}

{{}The $g_1^{\tau_2}(x, Q^2)$ and  $g_2^{\tau_2}(x, Q^2)$  structure functions at the leading twist order have valid definition  in the Bjorken limit, i.e. $Q^2 \rightarrow \infty,~x=$ fixed. But  at the a moderate low $Q^2$ ($\sim 1-5$ GeV$^2$) and $W^2 ($$4$ GeV$^2 <W^2<10$ GeV$^2$)  where $W^2$ is the invariant mass of the hadronic system, both TMC  along with  higher twist corrections should  be considered. We investigate them in Sec.\ref{target-mass-correction} and  Sec.\ref{Higher-twist-effect}.

Next section is devoted to illustrate the nucleon and deuteron structure functions, based on Jacobi  polynomial approach which  yield us these functions in moment-$n$ space.}

\section{Jacobi polynomials expansion technique\label{Jacobi-polynomials}}
{{} To achieve the nucleon structure function in momentum $n$-space we resort to a method that is based on the Jacobi polynomials expansion.}
Practical aspects of this method including its major advantages are presented in our previous studies~\cite{Shahri:2016uzl,Khanpour:2017cha,Khorramian:2010qa,Khorramian:2009xz,MoosaviNejad:2016ebo,Khanpour:2016uxh,AtashbarTehrani:2013qea,Nematollahi:2021ynm}.
According to this method, one can easily expand the polarized structure functions $x g_1^{\rm  QCD}(x,Q^2)$, in terms of the Jacobi polynomials $\Theta_{n}^{\alpha, \beta}(x)$, as it follows \cite{Ayala:2015epa,Barker:1982rv,Barker:1983iy,Krivokhizhin:1987rz,Krivokhizhin:1990ct,Chyla:1986eb,Barker:1980wu,Kataev:1997nc,
Alekhin:1998df,Kataev:1999bp,Kataev:2001kk,Kataev:2005ci,Leader:1997kw},
\begin{equation}\label{eq:xg1QCD}
	x \, g_{1}^{\tau 2}(x, Q^2) = x^{\beta} (1 - x)^{\alpha}\ \, \sum_{n = 0}^{\rm N_{ \rm max}} a_n(Q^2) \, \Theta_n^{\alpha, \beta}(x) \,,
\end{equation}
where $\rm N_{\rm max}$ is the maximum order of  expansion. The parameters $\alpha$ and $\beta$ are Jacobi polynomials free parameters which normally fixed on their best values.
These parameters have to be chosen so as to achieve the fastest convergence of the series on the right-hand side of Eq.~\eqref{eq:xg1QCD}.
In the polynomial fitting procedure, the evolution equation is combined with the truncated series to perform a direct fit to the structure functions.

{{} The Jacobi  moments, $a_{n}(Q^{2})$ are codifying the Q$^2$-dependence of the polarized structure functions.} The $x$-dependence will be provided by the weight function $w^{\alpha, \beta}(x) \equiv  x^{\beta} (1 - x)^{\alpha}$ and the Jacobi polynomials $\Theta_n^{\alpha, \beta}(x)$ which can be written as,
\begin{equation}\label{eq:Jacobi}
	\Theta_n^{\alpha, \beta}(x) = \sum_{j = 0}^{n} \, c_j^{(n)}(\alpha, \beta) \, x^j \,,
\end{equation}
where the coefficients $c_j^{(n)}(\alpha, \beta)$ are combinations of Gamma functions in terms of $n$, $\alpha$ and $\beta$. The above Jacobi polynomials {{}are satisfying the following  orthonormality condition:}
\begin{equation}\label{eq:orthogonality}
	\int_0^1 dx \, x^{\beta} (1 - x)^{\alpha} \, \Theta_n^{\alpha, \beta}(x) \, \Theta_{l}^{\alpha, \beta}(x) \, = \, \delta_{n, l} \,.
\end{equation}

{{}Consequently the Jacobi moments, $a_n(Q^2)$, can be obtained, using the above  relation such as,}
\begin{eqnarray} \label{eq:aMoment}
	a_n(Q^2) & = & \int_0^1 dx \, x g_1^{\tau_2}(x,Q^2) \, \Theta_n^{\alpha, \beta}(x)   \nonumber \\
	& = & \sum_{j = 0}^n \, c_j^{(n)}(\alpha, \beta) \, {\cal M} [xg_{1}^{\tau_2}, \, j + 2](Q^2)  \,,
\end{eqnarray}
where the Mellin transform ${\cal {M}} [x g_1^{\tau 2}, \rm N]$ {{} is given by},
\begin{eqnarray}\label{eq:Mellin}
	{\cal {M}} [x g_1^{\tau_2}, {\rm N}] (Q^2) \equiv  \int_0^1 dx \, x^{\rm N-2} \, xg_1^{\tau_2} (x, Q^2) \,.
\end{eqnarray}

{{} Using the QCD expressions for the Mellin moments, ${\cal {M}} [x g_1^{\tau_2}, {\rm N}] (Q^2)$,  the polarized structure function $x g_1^{\tau_2}(x, Q^2)$, can be constructed. Therefore, based on the method of Jacobi polynomial expansion, the $x g_1^{\tau_2}(x, Q^2)$ is given by:}
\begin{eqnarray}\label{eq:xg1Jacobi}
	x g_1^{\tau_2}(x, Q^2) & = & x^{\beta}(1 - x)^{\alpha} \, \sum_{n = 0}^{\rm N_{max}} \, \Theta_n^{\alpha, \beta}(x)    \nonumber   \\
	& \times & \sum_{j = 0}^n \, c_j^{(n)}{(\alpha, \beta)} \, {\cal M}[x g_1^{\tau_2}, j + 2](Q^2) \,.
\end{eqnarray}

{{} By setting  {$N_{\rm max}$ = 9, $\alpha$ = 3 and $\beta$ = 0.5}, as we have shown in our previous analyses \cite{Shahri:2016uzl,Khanpour:2017cha,Khorramian:2010qa,Khorramian:2009xz,MoosaviNejad:2016ebo,Khanpour:2016uxh,AtashbarTehrani:2013qea,Nematollahi:2021ynm}, it is possible to obtain the optimal convergence of above expansion through the whole kinematic region that is constrained by the polarized DIS data.

In next section we improve our analysis of DIS polarized data, considering the TMC correction to the nucleon structure functions.}
\section{Target mass corrections in polarized case}\label{target-mass-correction}

 Power suppressed corrections to the structure functions can have important contributions in some kinematic regions. Hence nucleon mass correction cannot be neglected in low $Q^2$ region.
The TMCs can be calculated via an expression which is different from  higher twist (HT) effects in dynamical case.
In the case of polarized structure function we follow the suggested method  by Blumlein and Tkabladze~ \cite{Blumlein:1998nv}  which is in fact the generalized  one that was established by Georgi and Politzer~\cite{Georgi:1976ve} for the unpolarized structure function.

Mellin inversion to $x$-space or the integer moments of structure function can be used  to present these corrections. Leading twist-2 expression for $g_1$, that is containing  {TMC, is given explicitly by \cite{Blumlein:1998nv}:
}
\begin{eqnarray}\label{eq:g1TMC}
	&& g_1^{\rm \tau_2 + TMCs}(x,Q^2)  \nonumber  \\
	& = & \frac{xg_1^{\tau_2}(\xi, Q^2; {\rm M} = 0)}{\xi(1 + 4 {\rm M}^2x^2/Q^2)^{3/2}}   \nonumber  \\
	& + & \frac{4{\rm M}^2x^2}{Q^2}\frac{(x + \xi)}{\xi(1 + 4 {\rm M}^2x^2/Q^2)^2}\int_{\xi}^{1}\frac{d\xi'}{\xi'} g_1^{\tau_2}(\xi', Q^2; {\rm M}=0)  \nonumber  \\
	& - & \frac{4 {\rm M}^2 x^2}{Q^2}\frac{(2 - 4 {\rm M}^2x^2/Q^2)}{2(1 + 4 {\rm M}^2x^2/Q^2)^{5/2}}  \nonumber  \\
	& \times & \int_{\xi}^1\frac{d \xi'}{\xi'} \int_{\xi'}^{1}\frac{d\xi''}{\xi''} g_1^{\tau_2} (\xi'', Q^2; {\rm M} = 0) \,.
\end{eqnarray}

 The twist-2 contribution for the $g_2$ structure function, including TMC is {similarly presented by \cite{Blumlein:1998nv}:}
\begin{eqnarray}\label{eq:g2TMC}
	&& g_2^{\tau_2+ \rm TMCs}(x,Q^2)  \nonumber  \\
	& = & -\frac{xg_1^{\tau_2}(\xi, Q^2; {\rm M} = 0)}{\xi(1 + 4 {\rm M}^2x^2/Q^2)^{3/2}}  \nonumber  \\
	& + & \frac{x(1 - 4 {\rm M}^2x\xi/Q^2)}{\xi(1 + 4 {\rm M}^2x^2/Q^2)^2}\int_{\xi}^{1}\frac{d\xi'}{\xi'} g_1^{\tau_2}(\xi', Q^2; {\rm M} = 0)  \nonumber  \\
	& + & \frac{3}{2}\frac{4 {\rm M}^2 x^2/Q^2}{(1 + 4 {\rm M}^2 x^2/Q^2)^{5/2}}  \nonumber  \\
	& \times & \int_{\xi}^1\frac{d \xi'}{\xi'}\int_{\xi'}^{1}\frac{d \xi''}{\xi''} g_1^{\tau_2}(\xi'', Q^2; {\rm M} = 0) \,,
\end{eqnarray}
{Numerical illustrations for the target mass effects in $g_1$ and $g_2$
have been given in~\cite{Blumlein:1999rv}.}
{{}In both above equations M is the nucleon mass and $\xi$ is called Nachtmann variable  that is defined by  \cite{Nachtmann:1973mr}:}
\begin{equation}
	\xi = \frac{2x}{1 + \sqrt{1 + 4 {\rm M}^2 x^2 / Q^2}}~.
\end{equation}

{{} It can be seen that by choosing the maximum value for the $x$-Bjorken variable, the maximum kinematic value of $\xi$ variable would be less than unity.
This means that the target mass corrected structure functions  at leading twist in both the polarized and unpolarized cases, as it is expected, do not vanish at maximum $x=1$ value.

As we referred before, in addition to target mass correction, higher twist effects would also be dominant at low $Q^2$  values and make contribution to nucleon structure function in related kinematic region. Next section is devoted to this effects.}

\section{Twist-3 contribution}\label{Higher-twist-effect}
{{} The long-range nonperturbative multiparton correlations which have outstanding contributions  at low values of $Q^2$ will lead to higher twist (HT) terms. {A proper analysis of this effect can be found in \cite{Blumlein:2010rn}.}
For a developing phenomenological analysis an advantageous parametrization is made by the BLMP model \cite{Braun:2011aw} for HT terms.
Following that HT distributions are constructed from convolution integrals that are containing light-cone wave functions. In this connection a simple model based on three valence quark and one gluon distributions with the total zero angular momentum are assumed.
}

Accordingly, we utilize the parameterized form, suggested by the BLMP model at the twist-3 order for $g_2$ structure function in an {initial scale $Q_0$  as it follows \cite{Braun:2011aw,Blumlein:2012se}}:
\begin{eqnarray}
	g_2^{{{\tau_3}}}(x)&=&A[ln(x)+(1-x)+\frac{1}{2}(1-x)^2]\nonumber\\
	&+&(1-x)^3[B-C(1-x)+D(1-x)^2\nonumber\\
	&-&E(1-x)^3]\ .
	\label{eq:g2HT}
\end{eqnarray}
{{}The unknown coefficients in Eq.(\ref{eq:g2HT}) are extracted by fitting the data. Since higher twist contributions are  important in a region with large-$x$ values, a nonsinglet evolution equation is employed. The results of this approach can be compared with exact evolution equations  where a gluon-quark-antiquark correlation is considered \cite{Braun:2011aw}. It is expecting that these two results are in good agreement with each other.

The twist-3 part of different spin-dependent structure functions , $g_1^{\tau_3}$ and $g_2^{\tau_3}$, are related by the following integral relation } \cite{Blumlein:1998nv}.
\begin{eqnarray}
	g_{1}^{\tau 3}(x,Q^2)&=&\frac{4x^2M^2}{Q^2}[g_2^{\tau 3}(x,Q^2)\nonumber\\
	&-&2\int_x^1 \frac{dy}{y}g_2^{\tau 3}(y,Q^2)]\ ,
	\label{eq:g1HT}
\end{eqnarray}
The $Q^2$-dependence of the $g_{\rm 2}^{\tau_3}$ can be achieved within nonsinglet perturbative QCD evolution as
\begin{equation}
	g_{\rm 2}^{\tau 3}(n, Q^2) = {\cal M}^{\rm NS}(n, Q^2) \,  g_{2}^{\tau 3}(n)\,.
\end{equation}
Finally the spin-dependent structure functions, considering the TMCs and  HT terms are given by,
\begin{eqnarray}\label{eq:g1full}
	x g_{1,2}&&^{\text {Full=pQCD+TMC+HT}}(x, Q^2) = \, \nonumber \\
	&&xg_{1,2}^{\rm \tau_2+TMCs}(x, Q^2) + xg_{1,2}^{\tau_3}(x, Q^2)\,.
\end{eqnarray}
%
{{} One of the particular feature of $x g_{1,2}^{\text {Full}}(x, Q^2)$ function is that the twist-3 term is not suppressed there by inverse powers of $Q^2$. Consequently to describe this function, this contribution  is  so important as the twist-2 contribution.

Since  the required theoretical inputs are accessed by us, we can do now the concerned data analysis which is done in next section}
\section{Fitting contents in QCD analysis}
The fitting procedure, including the recent and updated data for polarized structure functions which we do in our QCD analysis, are containing the following parts.
\label{QCD analysis and fitting procedure}
\subsection{Parametrization}
{{} We start  the QCD analysis considering} the following parametrization at the initial scale of $Q_{0}^{2}=1$ GeV$^{2}$  where  $q=\{u_{v},d_{v},\bar{q},g\}$:

\begin{equation}
	x\:\Delta q(x,Q_{0}^{2})={\cal N}_{q}\eta_{q}x^{a_{q}}(1-x)^{b_{q}}(1+c_{q}x)\ .
	\label{eq:parm}
\end{equation}
The normalization constant ${\cal N}_{q}$,
\begin{equation}
	{\cal N}_{q}^{-1}=\left(1+c_{q}\frac{a_{q}}{a_{q}+b_{q}+1}\right)\, B\left(a_{q},b_{q}+1\right)\ ,
	\label{eq:norm}
\end{equation}
is {{} determined}  such that $\eta_{q}$  in Eq.(\ref{eq:norm}) is the first moment of the {{} polarized parton distribution functions (PPDFs)}. Here
$B(a,b)$ is the Euler beta function.
Considering SU(3) flavor symmetry, we {{} assume} $\Delta\overline{q}\equiv\Delta\overline{u}=\Delta\overline{d}=\Delta s=\Delta\overline{s}$ .

{{} The unknown free parameters can be extracted through a fit which involves a large degree of flexibility. Some of parameters can be determined via the existing constrains, as describing in below:}

\begin{itemize}
	\item
{{} The  weak matrix elements $F$ and $D$ as measured
	in neutron and hyperon $\beta$ decays \cite{ParticleDataGroup:2018ovx} can be related to the first moments of the polarized valence quark densities. Considering these constrains, the numerical values $\eta_{u_{v}}=0.928\pm0.014$ and $\eta_{d_{v}}= -0.342\pm0.018$ are obtained.}

	\item
{{} Due to the present accuracy of the data, the $c_{\bar{q}}$ and $c_{g}$  parameters are setting to zero. Considering nonzero values for them, there would not be  observed any improvement in the fit.}

	\item
{{} The large-$x$ behavior of the polarized sea quarks and gluons are controlled  by $b_{\bar{q}}$ and $b_{g}$ parameters. In a region that is dominated by the valence distributions, these parameters have large uncertainties.}

 \item
{{} Due to higher twist effect to the  $g_{2,\{p,n,d\}}$ and consequently  $g_{1,\{p,n,d\}}$, there are unknown parameters $\{A,B,C,D,E\}$, sea Eq.(\ref{eq:g2HT}). By a simultaneous fit to the all polarized structure function data of $g_1$ and $g_2$, these parameters can be determined.}

\item
{{} The values of some parameters are frozen in the first minimization procedure.  They involve $\{\eta_{u_{v}},\eta_{d_{v}},c_{\bar{q}},c_{g}\}$ and finally the $b$ parameter.
As demonstrated in Tables~\ref{tab:result} and \ref{tab:g2twist3} the $\{b_{\bar{q}},b_{g},c_{u_{v}},c_{d_{v}}\}$  and $\{A,B,C,D,E\}$ parameters are then fixed in the  second minimization. Nine unknown parameters, including $\alpha_{s}(Q_{0}^{2})$,  are left which are determined in the fit. They have enough flexibility to perform a reliable fit.}

\item
{{} The numerical value $\alpha_{s}(M_{Z}^{2}) =0.112804\pm0.001907$  would be achieved in which  we need to change  the energy scale to the Z boson mass. It is while for the  present world average, the  value $\alpha_{s}(M_{Z}^{2}) = 0.1179\pm 8.5\times10^{-6}$ is reported ~\cite{Zyla:2020zbs}.

To extract the unknown parameters, it needs to access to  all available concerned data sets which we describe them in below.}
\end{itemize}

\begin{figure}[!htb]
	\includegraphics[clip,width=0.45\textwidth]{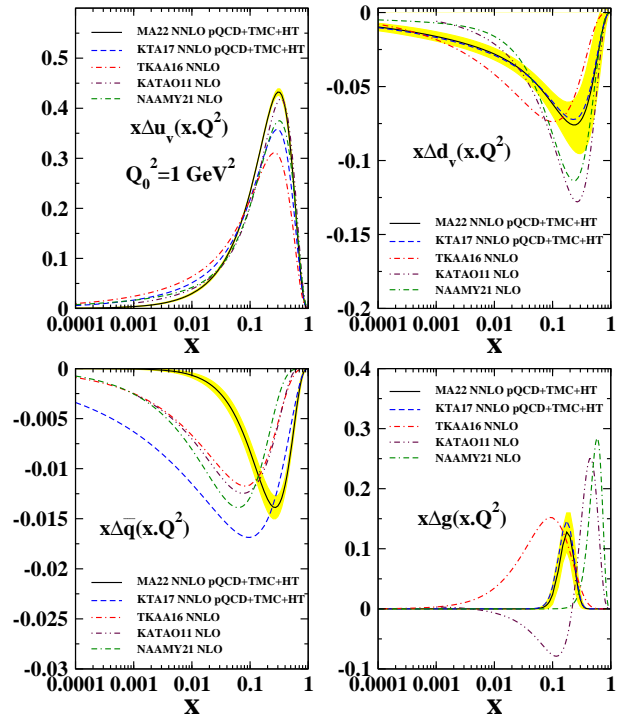}
	\begin{center}
		\caption{{\small Our MA22 results for the polarized PDFs at Q$_0^2=$ 1 GeV$^2$
				as a function of $x$ in the NNLO approximation. It is indicated by a solid curve along with their $\Delta \chi^2 = 1$
				uncertainty bands which is computed, based on the Hessian approach \cite{hes}.
				The recent results of TKAA16 (dashed-dotted)~\cite{Shahri:2016uzl} is also shown  in NNLO approximation without inclusion of HT terms and TMCs. Additionally the  KTA17(dashed)~\cite{Khanpour:2017cha} in NNLO approximation is presented that including the HT terms and TMCs. The KATAO11(dashed-dotted-dotted) in NLO approximation~\cite{Khorramian:2010qa} is furthermore indicated. Finally the results of NAAMY21(dashed-dashed-dotted)~\cite{Nematollahi:2021ynm} in NLO approximation is also plotted.  \label{fig:partonqcdhttmc}}}
	\end{center}
\end{figure}
\begin{table}
	\caption{\label{tab:result} Final parameter values and their statistical
		errors at the input scale $Q_0^2=1$ GeV$^2$ determined from two different global analyses. Those marked with ($^*$) are fixed.}
	\begin{tabular}{|c|c|c|c|}
		\hline\hline
		\multicolumn{2}{|c|}{Parameters}&{Full scenario}&{pQCD scenario}  \\
		\hline\hline $\delta u_{v}$ & $\eta_{u_{v}} $ & $~0.928^*~$& $~0.928^*~$  \\
		& $a_{u_{v}}$ & $0.898\pm 0.022$& $0.277\pm 0.0072$  \\
		& $b_{u_{v}}$ & $3.218\pm 0.035$& $2.725\pm 0.029$  \\
		& $c_{u_{v}}$ & $ 3.88^*$& $28.95^*$ \\  \hline
		$\delta d_{v}$ & $\eta_{d_{v}} $ & $-0.342^*$& $-0.342^*$  \\
		& $a_{d_{v}}$ & $0.217\pm 0.027$& $0.150\pm 0.012$ \\
		& $b_{d_{v}}$ & $2.947\pm 1.45$& $2.591\pm 0.087$ \\
		& $c_{d_{v}}$ & $~9.335^*~$& $~31.75^*~$  \\ \hline
		$~\delta_{\bar{q}}$ & $~\eta_{\bar{q}}$ & $-0.0288\pm 0.002$& $-0.0356\pm 0.0033$  \\
		& $a_{\bar{q}}$ & $1.227\pm 0.068$& $1.991\pm 0.041$   \\
		& $b_{\bar{q}}$ & $3.364^*$ & $11.163^*$   \\
		& $c_{\bar{q}}$ & $~0.0^*~$& $~0.0^*~$ \\ \hline
		$\delta g$ & $\eta_{g} $ & $0.0921\pm 0.022$& $0.178\pm 0.014$ \\
		& $a_{g}$ & $10.2\pm 1.22$& $26.33\pm 0.49$  \\
		& $b_{g}$ & $46.32^*$ & $99.95^*$ \\
		& $c_{g}$ & $~0.0^*~$& $~0.0^*~$  \\ 
		\hline
		\multicolumn{2}{|c|}{$\alpha _{s}(Q_{0}^2)$}&{$0.3362\pm 0.002$}&{$0.4688\pm 0.0008$} \\
		\hline \multicolumn{2}{|c|}{$\chi ^{2}/ndf$}&{$1111.789/957=1.161$}&{$1580.751/957=1.651$} \\
		\hline\hline
	\end{tabular}
\end{table}
%

\begin{table}[!htbp]
	\caption{\label{tab:g2twist3} Parameter values for the coefficients of the twist-3 corrections at $Q^2=1$ GeV$^2$ obtained in the full scenario.}
	\begin{ruledtabular}
		\begin{tabular}{lccccc}
			& \textbf{A}  & \textbf{B}  & \textbf{C}  &\textbf{D} &\textbf{E}  \\   \hline\hline
			$g_{2,p}^{tw-3}$  & $0.0879$ &$1.0196 $ & $-0.8832 $ &  $-2.3765$ &  $2.4234$ \\
			$g_{2,n}^{tw-3}$  & $1.0086$ &$0.3009 $ & $-0.6583$ &  $0.3466$ &  $-2.7571$ \\
			$g_{2,d}^{tw-3}$  & $0.8878$ &$1.3430 $ & $-2.1334 $ &  $0.1878$ &  $2.2293$ \\
		\end{tabular}
	\end{ruledtabular}
\end{table}
%
\begin{figure}[!htb]
	\includegraphics[clip,width=0.5\textwidth]{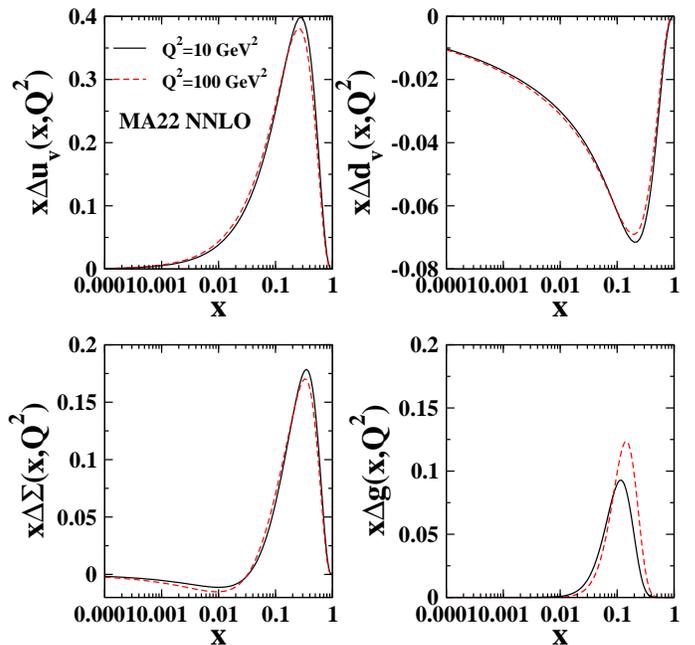}
	\begin{center}
		\caption{{\small  Our results,  MA22 polarized parton distributions as a function of $x$ and for some selected value of Q$^2$ = 10, 100 GeV$^2$. \label{fig:partonQ}}}
	\end{center}
\end{figure}
\subsection {Overview of data sets}
{{} In  our recent analysis which we call it MA22 we focus on the polarized DIS data samples.. The required DIS data for all PPDFs are coming from the experiments at electron-proton collider and also in fixed-target including proton, neutron and heavier targets such as deuteron.

 Although it is not possible to separate quarks from antiquarks, nonetheless it is the inclusive DIS data that are included in the fit. Additionally we take into  our MA22 fitting procedure the $g_2$ structure  function. Due to the technical difficulty in operating the required transversely polarized target, these data  have been traditionally neglected before. }

 {{} The data which we use in our recent analysis are up to date and including more data than we employed in our pervious analysis \cite{Khanpour:2017cha}. In fact} we use all available $g_1^p$ data from E143, HERMES98, SMC, EMC, E155, HERMES06, COMPASS10, COMPASS16, JLAB06 and JLAB17 experiments ~\cite{Abe:1998wq,HERM98,Adeva:1998vv,Ashman:1987hv,E155p,HERMpd,COMP1,Adolph:2015saz,Dharmawardane:2006zd,Fersch:2017qrq},  and $g_1^n$ data from HERMES98, E142, E154, HERMES06, Jlab03, Jlab04 and Jlab05~\cite{HERM98,E142n,E154n,Ackerstaff:1997ws,JLABn2003,JLABn2004,JLABn2005} and finally the $g_1^d$ data from E143, SMC, HERMES06, E155, COMPASS05, COMPASS06 and COMPASS17~\cite{Abe:1998wq,Adeva:1998vv,HERMpd,E155d,COMP2005,COMP2006,Adolph:2016myg}.
The DIS data for $g_2^{p, n, d}$ from E143, E142, Jlab03, Jlab04, Jlab05, E155, Hermes12 and SMC~\cite{Abe:1998wq,E142n,JLABn2003,JLABn2004,JLABn2005,E155pdg2,hermes2012g2,SMCpg2} also are included.
These data sets are summarized in Table~\ref{tab:DISdata}. The kinematic coverage, the number of data points for each given target, and the fitted normalization shifts ${\cal{N}}_i$ also presented in this Table.
{{} Our MA22 analysis algorithm computes the $Q^2$ evolution and extracts the structure function in $x$ space using Jacobi polynomials approach. It is corresponding to the fitting programs on the market which solve the DGLAP evolution equations in the Mellin space.

One of the important quantity which is used as a criteria to indicate the validation of fit, is  the chi-square ($\chi^2$) test which is assessing the goodness of fit between observed values and those expected theoretically. We discuss about it in the following subsection.}
\begin{figure}[!htb]
	\includegraphics[clip,width=0.5\textwidth]{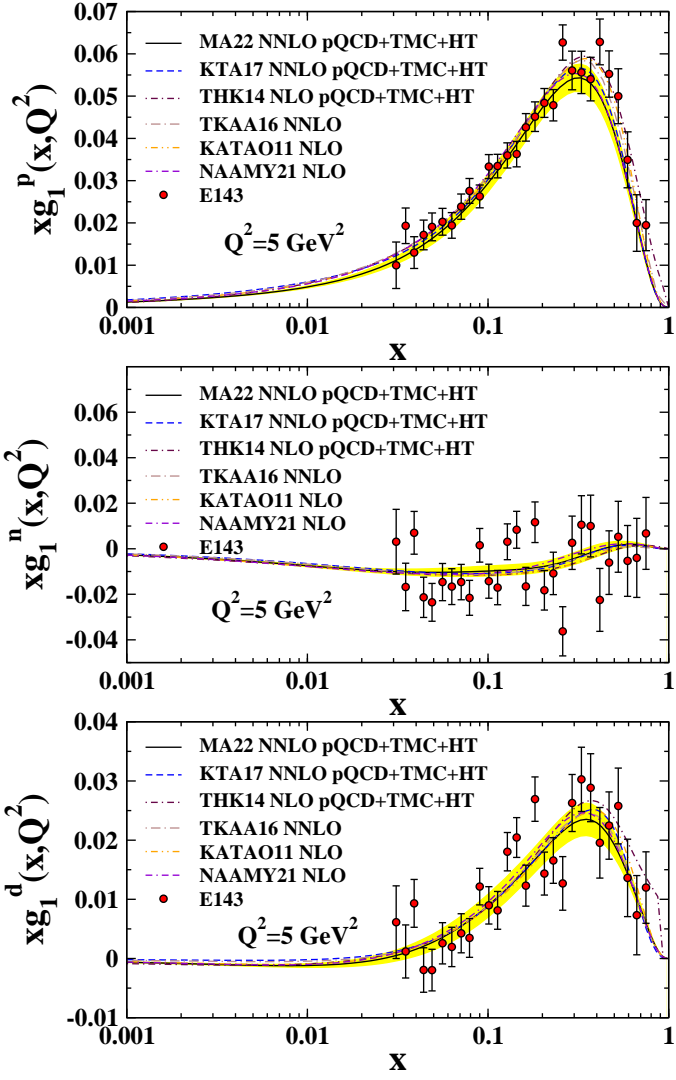}
	\begin{center}
		\caption{{\small   The spin-dependent proton, neutron and deuteron structure functions as a function of $x$ and Q$^2$. Our results, MA22, at the NNLO approximation (solid curve) are compared with KTA17 at the same approximation (dashed) ~\cite{Khanpour:2017cha}, with  THK14 at the NLO approximation (dashed-dotted)~\cite{TaheriMonfared:2014var}, with TKAA16  at the NNLO approximation (long-dashed dotted)~\cite{Shahri:2016uzl},with KATAO11 at the NLO approximation (dashed-dotted-dotted)~\cite{Khorramian:2010qa} and finally with  NAAMY21 at the NLO approximation (dashed-dashed-dotted)~\cite{Nematollahi:2021ynm}. \label{fig:xg1p}}}
	\end{center}
\end{figure}
\subsection{ $\chi^2$ minimization }
{{} The $\chi_{\rm global}^2(\rm p)$ quantifies the goodness of fit to the data for a set of  $\rm p$ independent parameters. To determine the best fit, it is needed to minimize the $\chi^2_{\rm global}$ function with the free unknown parameters. We do it for PPDFs at the NNLO approximation which  additionally includes  the QCD cut off parameter, $\Lambda_{\rm QCD}$ which finally specifies the polarized PDFs at Q$_0^2$ = 1 GeV$^2$.

This function is presented as it follows:}
%
\begin{equation}\label{eq:chi2}
	\chi_{\rm global}^2 ({\rm p}) = \sum_{n=1}^{N_{\rm exp}} w_n \chi_n^2\,.
\end{equation}
In this equation, ${w_n}$ is a weight factor for the $n^{\rm th}$ experiment and $\chi_n^2$ is defined by:
\begin{equation}\label{eq:chi2global}
	\chi_n^2 (\rm p) = \left( \frac{1 -{\cal N}_n }{\Delta{\cal N}_n}\right)^2 + \sum_{i=1}^{N_n^{\rm data}} \left(\frac{{\cal N}_n  \, g_{(1,2), i}^{\rm Exp} - g_{(1,2), i}^{\rm Theory} (p) }{{\cal N}_n \, \Delta g_{(1,2), i}^{\rm Exp}} \right)^2\,.
\end{equation}
%
The minimization of the above $\chi_{\mathrm{\rm global}}^2 (\rm p)$ function is done using the CERN program library MINUIT~\cite{James:1994vla}.
In the above equation, the main contribution comes from the difference between the model and the DIS data within the statistical precision.
{{} In the $\chi_n^2$ function, $g^{\rm Theory}$ indicates the theoretical value for the $i^{\rm th}$ data point  and  $g^{\rm Exp}$, $\Delta g^{\rm Exp}$ represent the experimental measurement and the experimental uncertainty (statistical and systematic combined in quadrature) respectively.

To do a proper fit an over normalization factor for the data of experiment $n$  is needed which is denoted by ${\cal N}_n$. An uncertainty ${\Delta{\cal N}_n}$ is attributed to this factor which should be considered in the fit. These factors, considering the uncertainties, quoted
by the experiments are used to relate  different experimental data sets. In fact they are taken as a free parameters which  are determined simultaneously with the other parameters in the fit. They are obtained in the  pre-fitting procedure and then fixed at their best values in further steps.
Numerical results for the unknown parameters, resulted from $\chi^2$  minimization, are listed in Table.\ref{tab:result} and \ref{tab:g2twist3}. Different data sets which are  used in the fit, is presented in Table.\ref{tab:DISdata}.

Now we are at stage to do some analytical computations for a more  confirmation of  the fitting validation, taken into account the several sun rules as we do it in the next section.}

\begin{table*}[htb]
	\caption{Summary of published polarized DIS experimental data points with measured $x$ and $Q^2$ ranges and the number of data points.} \label{tab:DISdata}
	\begin{tabular}{l c c c c c c}
		\hline
		\textbf{Experiment} & \textbf{Ref.} & \textbf{[$x_{\rm min}, x_{\rm max}$]}  & \textbf{Q$^2$  {(}GeV$^2${)}}  & \textbf{Num. of data poi.} & $\chi^2$ & ${\cal N}_i$     \tabularnewline
		\hline\hline
		\textbf {SLAC/E143(p)}   & \cite{Abe:1998wq}   & [0.031--0.749]   & 1.27--9.52 & 28& 19.0218 &0.99705 \\
		\textbf{HERMES(p)} & \cite{HERM98}  & [0.028--0.66]    & 1.01--7.36 & 39& 55.2816 &0.99982 \\
		\textbf{SMC(p)}    & \cite{Adeva:1998vv}    & [0.005--0.480]   & 1.30--58.0 & 12& 8.9328 &1.00009 \\
		\textbf{EMC(p)}    & \cite{Ashman:1987hv}     & [0.015--0.466]   & 3.50--29.5 & 10& 3.8416 &1.00592\\
		\textbf{SLAC/E155}      & \cite{E155p}    & [0.015--0.750]   & 1.22--34.72 & 24& 41.7453 &0.99915 \\
		\textbf{HERMES06(p)} & \cite{HERMpd} & [0.026--0.731]   & 1.12--14.29 & 51& 21.0559 &0.99915 \\
		\textbf{COMPASS10(p)} & \cite{COMP1} & [0.005--0.568]   & 1.10--62.10 & 15& 23.1003 &1.00073\\
		\textbf{COMPASS16(p)} & \cite{Adolph:2015saz} & [0.0035--0.575]   & 1.03--96.1 & 54& 52.6140 &1.00296 \\
		\textbf {SLAC/E143(p)}   & \cite{Abe:1998wq}   & [0.031--0.749]   & 2-3-5 & 84& 122.0060 &0.99578 \\
		\textbf{HERMES(p)} & \cite{HERM98}  & [0.023--0.66]    &      2.5    & 20& 35.2073 &0.99726 \\
		\textbf{SMC(p)}    & \cite{Adeva:1998vv}    & [0.003--0.4]   & 10 & 12& 14.8138 &1.00071 \\
		\textbf{Jlab06(p)}&\cite{Dharmawardane:2006zd} & [0.3771--0.9086]     &3.48--4.96 & 70& 99.6438 &1.00127\\
		\textbf{Jlab17(p)}&\cite{Fersch:2017qrq} & [0.37696--0.94585]     &3.01503--5.75676     & 82&  171.5716 & 1.00282\\
		\multicolumn{1}{c}{$\bf{g_1^p}$}         &  &  &   &  \textbf{501} && \\
		\textbf{SLAC/E143(d)}  &\cite{Abe:1998wq}    & [0.031--0.749]   & 1.27--9.52    & 28& 38.3735 &1.00210\\
		\textbf{SLAC/E155(d)}  &\cite{E155d}     & [0.015--0.750]   & 1.22--34.79   & 24& 20.0319 &1.00228 \\
		\textbf{SMC(d)}   &\cite{Adeva:1998vv}     & [0.005--0.479]   & 1.30--54.80   & 12& 18.3574 &1.00006 \\
		\textbf{HERMES06(d)} & \cite{HERMpd}& [0.026--0.731]   & 1.12--14.29   & 51& 44.4642 &1.00654  \\
		\textbf{COMPASS05(d)}& \cite{COMP2005}& [0.0051--0.4740] & 1.18--47.5   & 11& 7.3430 &1.00760 \\
		\textbf{COMPASS06(d)}& \cite{COMP2006}& [0.0046--0.566] & 1.10--55.3    & 15& 8.4408 &1.00052 \\
		\textbf{COMPASS17(d)} & \cite{Adolph:2016myg} & [0.0045--0.569]   & 1.03--74.1 & 43& 36.2019 &1.01090  \\
		\textbf{SLAC/E143(d)}  &\cite{Abe:1998wq}    & [0.031--0.749]   & 2--3--5    & 84& 127.5502 &0.99981\\
		\multicolumn{1}{c}{ $\bf{g_1^d}$}        &  & & & \textbf{268}  & &   \\
		\textbf{SLAC/E142(n)}   &\cite{E142n}    & [0.035--0.466]   & 1.10--5.50    & 8& 8.0235 &0.99881\\
		\textbf{HERMES(n)} &\cite{HERM98}   & [0.033--0.464]   & 1.22--5.25    & 9& 2.7585 &0.99995 \\
		\textbf{E154(n)}   &\cite{E154n}    & [0.017--0.564]   & 1.20--15.00   & 17& 14.6888 &0.99908 \\
		\textbf{HERMES06(n)} &\cite{Ackerstaff:1997ws}  &  [0.026--0.731]  & 1.12--14.29   & 51& 18.1873 &0.99913 \\
		\textbf{Jlab03(n)}&\cite{JLABn2003} & [0.14--0.22]     & 1.09--1.46    & 4&1.803e-2 &0.99950 \\
		\textbf{Jlab04(n)}&\cite{JLABn2004} & [0.33--0.60]      & 2.71--4.8     & 3&2.2174 &1.05642\\
		\textbf{Jlab05(n)}&\cite{JLABn2005} & [0.19--0.20]     &1.13--1.34     & 2&3.2639 &0.98666\\
		\multicolumn{1}{c}{$\bf{g_1^n}$}     &  &     & & \textbf{94} & &  \\
		\textbf{E143(p)}    & \cite{Abe:1998wq}   & [0.038--0.595]   & 1.49--8.85    & 12&7.1338&1.00074 \\
		\textbf{E155(p)}   &\cite{E155pdg2}  &[0.038--0.780]    & 1.1--8.4      & 8 &11.9908&0.99886 \\
		\textbf{Hermes12(p)}&\cite{hermes2012g2} &[0.039--0.678]&1.09--10.35   & 20&22.6010&0.99898 \\
		\textbf{SMC(p)}      &\cite{SMCpg2} & [0.010--0.378]    & 1.36--17.07   & 6 &1.6804&1.00000 \\
		\multicolumn{1}{c}{$\boldsymbol{g_2^p}$}  &  &  & & \textbf{46} &    \\
		\textbf{E143(d)}     &\cite{Abe:1998wq} & [0.038--0.595]   & 1.49--8.86    & 12 &8.3504&1.00010 \\
		\textbf{E155(d)}    &\cite{E155pdg2}& [0.038--0.780]    & 1.1--8.2      & 8 &1.9800&1.00296 \\
		\multicolumn{1}{c}{$\boldsymbol{g_2^d}$}  &  &  & & \textbf{20} &   \\
		\textbf{E143(n)}    &\cite{Abe:1998wq}  & [0.038--0.595]   & 1.49--8.86    & 12&8.87903&1.00001 \\
		\textbf{E155(n)}    &\cite{E155pdg2}&[0.038--0.780]    &1.1--8.8       & 8 &6.0324&1.01893 \\
		\textbf{E142(n)}    &\cite{E142n}   &[0.036--0.466]    &1.1--5.5       & 8 &3.8955&0.99999 \\
		\textbf{Jlab03(n)}  &\cite{JLABn2003}&[0.14--0.22]     & 1.09--1.46    & 4 &0.9362&0.99337 \\
		\textbf{Jlab04(n)}  &\cite{JLABn2004}&[0.33--0.60]     & 2.71--4.83    & 3 &3.9915&1.10299 \\
		\textbf{Jlab05(n)}  &\cite{JLABn2005}&[0.19--0.20]     & 1.13--1.34    & 2 &15.5600&0.98986 \\
		\multicolumn{1}{c}{$\boldsymbol{g_2^n}$}  &  &  & &\textbf{37} &   \\  \hline
		\hline\\
		\multicolumn{1}{c}{\textbf{ Total}}&\multicolumn{4}{c}{~~~~~~~~~~~~~~~~~~~~~~~~~~~~~~~~~~~~~~~~~~~~~~~~~~~~~~~\textbf{966}}&\multicolumn{1}{c}{\textbf{1111.7891}}
		\\
		\hline\\
	\end{tabular}
\end{table*}

\begin{figure}[!htb]
	\begin{center}
	\includegraphics[clip,width=0.6\textwidth]{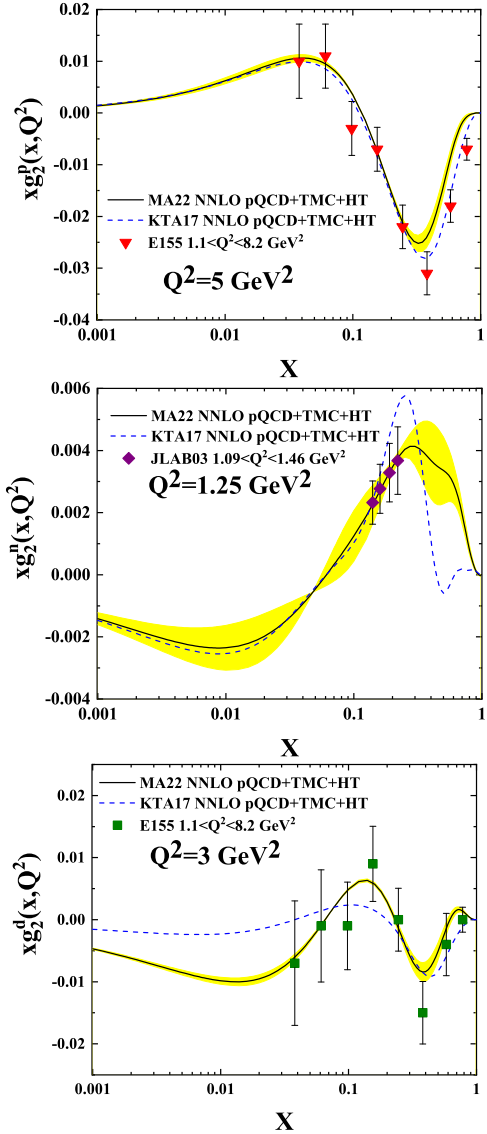}
		\caption{{\small  The spin-dependent proton, neutron and deuteron structure functions, $xg_2$, as a function of $x$ and Q$^2$. Our results, MA22, at the NNLO approximation (solid curve) are compared with KTA17  at the same approximation (dashed)~\cite{Khanpour:2017cha}. ~ \label{fig:xg2p}}}
	\end{center}
\end{figure}
%
\begin{figure}[!htb]
	\begin{center}
		\includegraphics[clip,width=0.5\textwidth]{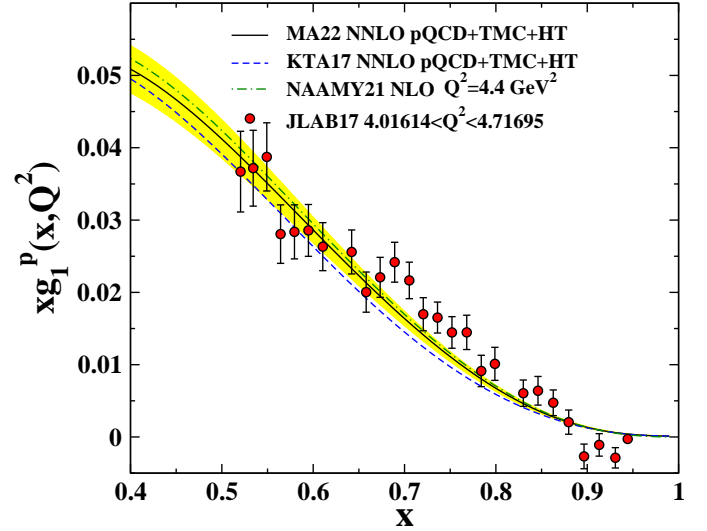}
		\caption{{\small   The spin-dependent proton structure function, $xg_1$, as a function of $x$ and Q$^2$. Our result, MA22,  at the NNLO approximation (solid curve) is compared with KTA17  at the same approximation (dashed) ~\cite{Khanpour:2017cha}  and with NAAMY21 at the NLO approximation (dashed-dotted)~\cite{Nematollahi:2021ynm}.~ \label{fig:xg1pjlab}}}
	\end{center}
\end{figure}

\section{The Sum Rules} \label{Sum-Rules}
{{} Sum rules like total momentum fraction carried by partons or the total contribution of parton spin to the spin of the nucleon are important tools to investigate some fundamental properties of the nucleon structure. Inclusion of TMCs and HT terms into the NNLO polarized structure function analysis leads to an improvement for the precision of PPDF determination as well as QCD sum rules  and we are exploring herein their effects. In what are following by utilizing available experimental data, we describe some important polarized sum rules.}

\subsection{Bjorken sum rule} \label{Bjorken-sum-rule}
{{} The polarized Bjorken sum rule expresses the integral over the spin distributions of quarks inside  the nucleon in terms of its axial charge, $g_A$ (as measured in neutron $\beta$ decay), times a coefficient function, $C_{Bj}[\alpha_s(Q^2)]$ ~\cite{Bjorken:1969mm}, and  considering higher twist (HT) corrections, it is given by}
\begin{eqnarray}\label{eq:Bjorken-SR}
	\Gamma_1^{\rm NS}(Q^2)&=&\Gamma_1^p(Q^2) - \Gamma_1^n(Q^2)   \nonumber \\
	&=& \int_0^1[g_{1}^{p}(x, Q^2) - g_1^{{n}}(x, Q^2)]dx        \nonumber \\
	&=& \frac{1}{6} ~ |g_A|~ C_{Bj}[\alpha_s(Q^2)] + \text{HT corrections}\,.\nonumber \\
\end{eqnarray}
{{} Bjorken sum rule potentially provides a very precise handle on the $\alpha_s$ as strong coupling constant. The value of  coupling can be extracted via $C_{Bj}[\alpha_s(Q^2)]$ expression from experimental data. This function has been calculated in 4-loop pQCD corrections in the massless \cite{Baikov:2010je} and very recently massive cases \cite{Blumlein:2016xcy}.
As previously reported in Ref.~\cite{Altarelli:1998nb}, determination of $\alpha_s$ from the Bjorken sum rule suffers from small-$x$ extrapolation ambiguities.

The $\alpha_s$ is also available form accurate methods to compute the width decay of $\tau$-lepton and the $Z$-boson into hadrons \cite{akrami1,akrami2}. An important test of QCD consistency can be offered by comparing these values.

Our results for the Bjorken sum rule can be compared with experimental measurements such as \text{E143}~\cite{Abe:1998wq}, \text{SMC}~\cite{SMCpg2}, \text{HERMES06}~\cite{HERMpd} and \text{COMPASS16}~\cite{Adolph:2015saz}.  The comparisons indicate an adequate consistency as we list them in Table~\ref{tab:Bjorken}.}
%
%
\begin{table*}[!htb]
	\caption{\label{tab:Bjorken} Comparison  our computed MA22 result for the Bjorken sum rule, $\Gamma_1^{NS}$, with world data from
		\text{E143}~\cite{Abe:1998wq}, \text{SMC}~\cite{SMCpg2}, \text{HERMES06}~\cite{HERMpd} and \text{COMPASS16}~\cite{Adolph:2015saz}.
		Only HERMES06~\cite{HERMpd} results are not extrapolated in full $x$ range (measured in region $0.021 \leq x \leq 0.9$). }
	\begin{ruledtabular}
		\begin{tabular}{lcccccc}
			& \textbf{E143}~\cite{Abe:1998wq}   & \textbf{SMC}~\cite{SMCpg2}  & \textbf{HERMES06}~\cite{HERMpd}  &  \textbf{COMPASS16}~\cite{Adolph:2015saz} & \textbf{KTA17}~\cite{Khanpour:2017cha} & \textbf{MA22}\\ 
			& $Q^2=5$ GeV$^2$& $Q^2=5$ GeV$^2$& $Q^2=5$ GeV$^2$& $Q^2=3$ GeV$^2$&$Q^2=5$ GeV$^2$  &$Q^2=5$ GeV$^2$     \\     \hline    \hline\tabularnewline
					$\Gamma^{\rm NS}_1$   & $0.164 \pm 0.021$ & $ 0.181\pm 0.035 $  & $0.148 \pm 0.017$ & $0.181\pm 0.008$& $0.167\pm0.005$ & $0.171\pm0.001$    \\
		\end{tabular}
	\end{ruledtabular}
\end{table*}
\begin{figure}[!htb]
	\begin{center}
		\includegraphics[clip,width=0.5\textwidth]{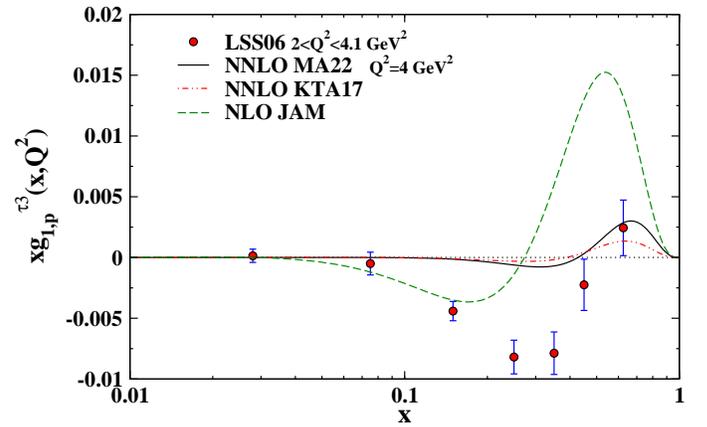}
		\caption{{\small   The twist-3 contribution to $xg_1^p$ at $Q^2$=4 GeV$^2$ as a function of $x$ is compared with the results of LSS~\cite{Leader:2006xc} and KTA17~\cite{Khanpour:2017cha} at NNLO approximation  and with JAM~\cite{Jimenez-Delgado:2013boa} at the NLO approximation. \label{fig:xg1t3Q1}}}
	\end{center}
\end{figure}
\begin{figure}[!htb]
	\begin{center}
		\includegraphics[clip,width=0.5\textwidth]{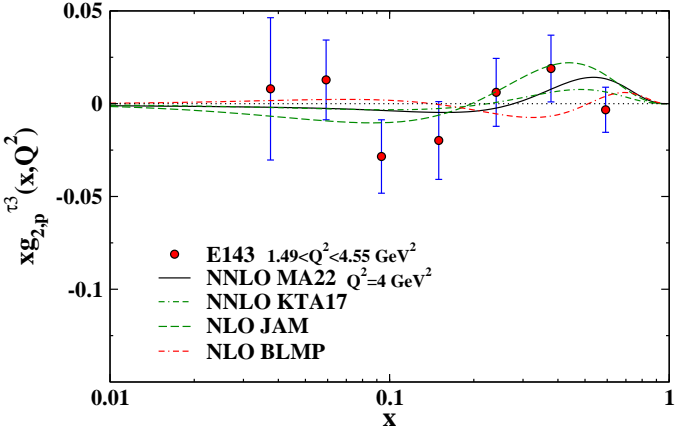}
		\caption{{\small  The twist-3 contribution to $xg_2^p$ at $Q^2$ = 4 GeV$^2$ as a function of $x$.  Our result, MA22 (solid curve), is compared with KTA17 at the NNLO approimation~\cite{Khanpour:2017cha}(dashed-dotted), JAM  ~\cite{Jimenez-Delgado:2013boa}(dashed) and  BLMP \cite{Braun:2011aw}
				(dashed dashed dotted) at the NLO approximation.  E143 experimental data~\cite{Abe:1998wq} have also been added. \label{fig:xg2twQ1}}}
	\end{center}
\end{figure}
\begin{figure}[!htb]
	\begin{center}
		\includegraphics[clip,width=0.39\textwidth]{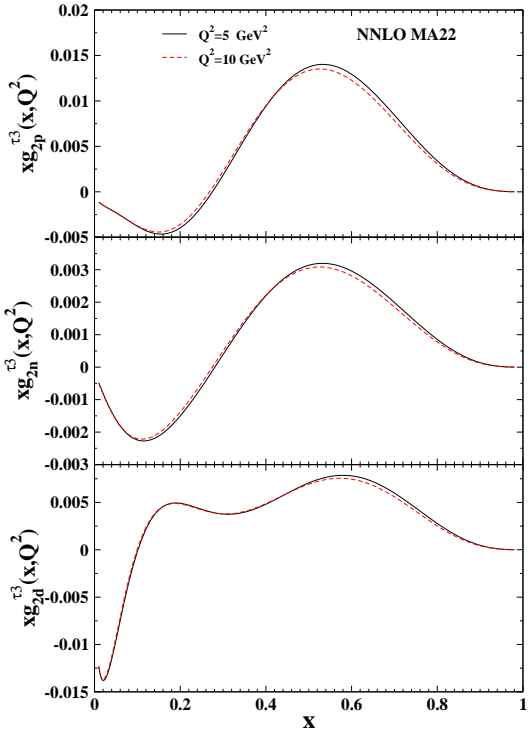}
		\caption{{\small The twist-3 contribution of $xg_2$ for the proton, neutron, and deuteron as a function of $x$ and for different
				values of Q$^2$ according to our result, MA22, at the NNLO analysis. \label{fig:xg2tw3}}}
	\end{center}
\end{figure}
\begin{figure}[!htb]
	\begin{center}
		\includegraphics[clip,width=0.39\textwidth]{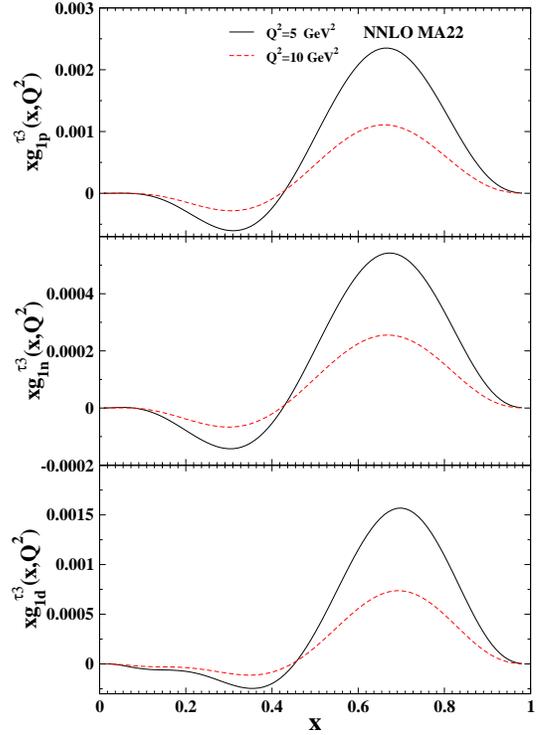}
		\caption{{\small   The twist-3 contribution of $xg_1$ for the proton, neutron, and deuteron as a function of $x$ and for different
				values of Q$^2$ according to our results,  MA22, at the NNLO analysis. \label{fig:xg1tw3}}}
	\end{center}
\end{figure}
\subsection{Proton helicity sum rule}
{{} This sum rule is related to the extrapolation of proton spin among its constituents that is completing our knowledge in the field of nuclear physics \cite{Leader:2016sli}.  An accurate picture of the quark and gluon helicity density are obtained, considering proton's momentum sum rule that needs a precise extraction of PPDFs.

The spin of the nucleon are carried by its constituents  that is generally represented by}
\begin{equation}\label{eq:spinsumrule}
	\frac{1}{2} = \frac{1}{2} \Delta \Sigma(Q^2) + \Delta {\mathrm G}(Q^2) + {\mathrm L}(Q^2).
\end{equation}
{{} Here $\Delta \Sigma(Q^2)=\sum_{i}\int_{0}^{1}dx~(\Delta q(x,Q^2)+\Delta \bar{q}(x,Q^2))$ denotes spin contribution of the singlet flavour, $\Delta {\rm G(Q^2)}=\int_{0}^{1}dx~\Delta g(x,Q^2)$  is interpreted as the gluon spin contribution and finally ${\mathrm L}(Q^2)$ represents the total contribution from quark and gluon orbital angular momentum. Each individual term in  Eq.(\ref{eq:spinsumrule})  is a function of $Q^2$ but the sum is not. Finding a way to measure them is a real challenge. Describing the measurement methods is the  beyond the scope of this paper.}

{{} In Table~\ref{tab:firstmomentQ10} the amount of first moment for  the singlet-quark and gluon are listed at Q$^2$=10 GeV$^2$.
Our results are compared to those from the \text{NNPDFpol1.0}~\cite{Ball:2013lla}, \text{NNPDFpol1.1}~\cite{Nocera:2012hx} and \text{DSSV08}~\cite{deFlorian:2008mr} at both truncated and full $x$ region.

In Table~\ref{tab:firstmomentQ4} our results, MA22, are presented and compared  with the results of \text{DSSV08}~\cite{deFlorian:2008mr}, \text{BB10}~\cite{Blumlein:2010rn}, \text{LSS10}~\cite{Leader:2010rb} , \text{NNPDFpol1.0}~\cite{,Ball:2013lla} and \text{KTA17}~\cite{Khanpour:2017cha}  at Q$^2$=4 GeV$^2$.}

{{} As can be seen from the Table~\ref{tab:firstmomentQ10} and Table~\ref{tab:firstmomentQ4}  for the $\Delta \Sigma$, our MA22 results are consistent within uncertainties with that of other groups. It is back to this reason that the first moment of polarized densities are mainly fixed by semileptonic decays. Very different values are reported by various groups when we turn to the gluon. Considering their large uncertainty are avoiding us to reach a firm conclusion about the full first moment of gluon.}

%
\begin{table*}[!htb]
	\caption{\label{tab:firstmomentQ10} Results for the full and truncated first moments of the polarized singlet-quark $\Delta \Sigma(Q^2)=\sum_{i}\int_0^1 dx [\Delta q_i(x) + \Delta \bar q_i (x)]$ and gluon distributions at the scale Q$^2$=10 GeV$^2$ in the $\overline{{\rm MS}}$--scheme. Also shown are the recent polarized global analysis of \text{NNPDFpol1.0}~\cite{,Ball:2013lla}, \text{NNPDFpol1.1}~\cite{Nocera:2012hx} and \text{DSSV08}~\cite{deFlorian:2008mr}. }
	\begin{ruledtabular}
		\begin{tabular}{lccccc}
			& \textbf{DSSV08}~\cite{deFlorian:2008mr}  & \textbf{NNPDFpol1.0}~\cite{,Ball:2013lla} & \textbf{NNPDFpol1.1}~\cite{Nocera:2012hx}&  \textbf{KTA17}~\cite{Khanpour:2017cha}  &  \textbf{MA22}   \tabularnewline
			Full $x$ region $[0,1]$ &  &  & & & \\
			\hline \hline \tabularnewline
			$\Delta \Sigma {\rm(Q^2)}$&$0.242$  &$+0.16\pm 0.30$ &$+0.18\pm 0.21$       &$0.2587\pm0.044$  &$0.2445\pm0.0048$  \\
			$\Delta {\rm G(Q^2)}$   & $-0.084$ & $-0.95\pm 3.87$ & $0.03\pm 3.24$     &$0.2104\pm0.034$ &$0.1205\pm0.03$ \\ \hline  \hline  \tabularnewline
			Truncated $x$ region [$10^{-3},1$] &  &  & & & \\     \hline \hline \tabularnewline
			$\Delta \Sigma {\rm(Q^2)}$  & $0.366\pm0.017$& $+0.23\pm0.15$&$+0.25\pm0.10$    &$0.2661\pm0.038$ &$0.2551\pm0.0066$\\
			$\Delta {\rm G(Q^2)}$   &$0.013\pm0.182$  & $-0.06\pm 1.12$ & $0.49\pm 0.75$  &$0.2104\pm0.034$ & $0.1205\pm0.03$
		\end{tabular}
	\end{ruledtabular}
\end{table*}
%
%
\begin{table*}[!htb]
	\caption{\label{tab:firstmomentQ4}Same as Table \ref{tab:firstmomentQ10}, but only for the full first moments of the polarized singlet-quark and gluon distributions at the scale Q$_0^2$ = 4 GeV$^2$ in the $\overline{{\rm MS}}$--scheme. Those of \text{DSSV08}~\cite{deFlorian:2008mr}, \text{BB10}~\cite{Blumlein:2010rn}, \text{LSS10}~\cite{Leader:2010rb} and \text{NNPDFpol1.0}~\cite{,Ball:2013lla} are presented for comparison. }
	\begin{ruledtabular}
		\begin{tabular}{ccccccc}
			& \textbf{DSSV08}~\cite{deFlorian:2008mr}  & \textbf{BB10}~\cite{Blumlein:2010rn} & \textbf{LSS10}~\cite{Leader:2010rb} & \textbf{NNPDFpol1.0}~\cite{,Ball:2013lla} &  \textbf{KTA17}~\cite{Khanpour:2017cha} &  \textbf{MA22}    \tabularnewline
			\hline\hline
			$\Delta \Sigma {\rm(Q^2)}$  & $0.245$ & $0.193 \pm  0.075$ & $0.207 \pm 0.034$ & $0.18 \pm 0.20$ & $0.1774\pm0.029$ & $0.2607\pm0.0065$ \\
			$\Delta {\rm G(Q^2)}$   & $-0.096$ & $0.462 \pm 0.430$ & $0.316 \pm 0.190$ & $-0.9 \pm 4.2$ & $0.1882\pm0.0294$ & $0.1095\pm0.027$  \\
		\end{tabular}
	\end{ruledtabular}
\end{table*}

{{} Based on the extracted values presented in Table~\ref{tab:firstmomentQ4} we can finally discuss the proton spin sum rule. Hence the amount of  quark and gluon orbital angular momentum to the  spin of the proton would be:}

\begin{equation}
	{\mathrm L}(Q^2=4~\rm GeV^2) = 
	0.3591 \pm 0.0779 \,.
\end{equation}

{{} A definite conclusion about the contribution of the total orbital angular momentum to the spin of the proton can not be done because of the large uncertainty that is  mainly  originating from the gluons. To obtain a precise determination of each individual contribution, it is required  to improve the current level of experimental accuracy.}

\subsection{The twist-3 reduced matrix element $d_2$}
{{} One of the quantity which is not considered as a sum rule but its numerical evaluation is remarkable to invastigate the higher twist effect is the twist-3 reduced matrix element and is denoted by  $d_2$. {Detailed of higher twist analyses for $g_1$ polarized structure function have been performed in  \cite{Blumlein:2010rn}.} In operator product expansion (OPE) theorem  \cite{ope} the effect of quark-gluon correlations can be studied through the moments of $g_1$ and $g_2$ structure functions. These moments lead to definition of reduced matrix element, $d_2(Q^2)$, as it follows}
\begin{eqnarray}
	d_2(Q^2) &=& 3 \int_0^1 x^2 \bar{g_2}(x,Q^2)~dx       \nonumber   \\
	&=& \int_0^1 x^2 [3 g_2(x, Q^2) + 2g_1(x, Q^2)]~ dx .
\label{d2}
\end{eqnarray}
{{} In this equation $\bar g_2=g_2-g_{2} ^{WW}$ where $g_{2} ^{WW}$ is given by Wandzura and
Wilczek (WW) relation as in Eq.(\ref{eq:WW}). The $d_2(Q^2)$ that  is in fact the twist-3 reduced matrix element of spin dependent operators in nucleon, can be used to measure the deviation of $g_2$ from $g_{2} ^{\tau_2}$. Due to the $x^2$ weighting factor in Eq.(\ref{d2}), this matrix element  is specially sensitive to the large-$x$ behaviour of $\bar{g_2}$. Some insights into the size of the multi-parton correlation terms can be obtained by extracting  the $d_2$  which indicates its important.

The significance of higher twist terms in QCD analyses is revealed by having non-zero value for $d_2$. To achieve precise information on the higher twist operators and to improve model prediction, a much more accurate experimental measurement for $d_2$ is required. In Table~\ref{tab:twist3} we present our results for $d_2$ which are compared with the other theoretical predictions  and also experimental values.}
%
%
%
\begin{table*}[!htb]
	\caption{\label{tab:twist3} $d_2$ moments of the proton, neutron and deuteron polarized structure functions from the \text{SLAC E155x}~\cite{Kuhn:2008sy},
		\text{E01-012}~\cite{Solvignon:2013yun}, \text{E06-014}~\cite{Flay:2016wie}, \text{Lattice QCD}~\cite{Gockeler:2005vw}, \text{CM bag model}~\cite{Song:1996ea}, \text{JAM15}~\cite{Sato:2016tuz}, \text{JAM13}~\cite{Jimenez-Delgado:2013boa}, KTA17~\cite{Khanpour:2017cha} compared with MA22 results.
	}
	\begin{ruledtabular}
		\begin{tabular}{lccccc}
			& Ref. &  \textbf{$Q^2$ [GeV$^{2}$]} &  \textbf{$10^2d^{p}_2$}  & \textbf{$10^5d^{n}_2$}  & \textbf{$10^3d^d_2$} \tabularnewline  \hline  \hline  \tabularnewline
			MA22      &&$5$&$1.0929\pm0.0106$         & $209.095\pm3.96$        & $7.206\pm0.078$       \\
			KTA17      &\cite{Khanpour:2017cha}&$5$&$0.718\pm0.01$         & $105.36\pm74.58$        & $5.16\pm0.02$       \\
			\text{E06-014} & \cite{Flay:2016wie}& 3.21 & &$-421.0 \pm 79.0 \pm 82.0 \pm 8.0$ & -\\
			\text{E06-014} & \cite{Flay:2016wie}& 4.32 & &$-35.0 \pm 83.0 \pm 69.0 \pm 7.0$  & -\\
			\text{E01-012} & \cite{Solvignon:2013yun} &3 & - & $-117 \pm 88 \pm 138$ &  -  \\
			\text{E155x}   & \cite{E155pdg2} & $5$ & $0.32\pm 0.17$ & $790\pm 480$ &     -                         \\
			\text{E143}    & \cite{Abe:1998wq}& $5$                 & $0.58\pm 0.50$ & $500\pm 2100$ & $5.1\pm 9.2$ \\
			\text{Lattice QCD}  &\cite{Gockeler:2005vw} & 5 &  0.4(5) & -100(-300) & - \\
			\text{CM bag model} &\cite{Song:1996ea} & $5$  & $1.74$          & $-253$       & $6.79 $         \\
			\text{JAM15}        &\cite{Sato:2016tuz} & $1$  & $0.5\pm 0.2$          & $-100 \pm 100$       & -        \\
			\text{JAM13}        &\cite{Jimenez-Delgado:2013boa} & $5$  & $1.1 \pm 0.2$          & $200 \pm 300$       &  -         \\
			
		\end{tabular}
	\end{ruledtabular}
\end{table*}
%

%
\subsection{Burkhardt-Cottingham (BC) sum rule} \label{BC-sum-rule}
{{} Considering dispersion relations for virtual Compton scattering in all Q$^2$, Burkhardt and Cottingham predicted that the zeroth moment of $g_2$ goes to zero ~\cite{Burkhardt:1970ti} such as: }

\begin{equation}
	\Gamma_2 = \int_0^1 dx \, g_2(x, Q^2) = 0~.
\end{equation}
{{} This relation is called Burkhardt-Cottingham (BC) sum rule and is  trivial consequence of the WW relation for $g_{2}^{\tau_2}$  (see Eq.(\ref{eq:WW})). {It should be noted that zeroth moment of structure function does not exist in the  light cone expansion and hence can not be described by local operator product expansion \cite{Blumlein:1996vs}.} Even if the target mass corrected structure function is used, this sum rule is still established \cite{Blumlein:1998nv}. Consequently any violation of the BC sum rule is an evidence for the presence of HT contributions~\cite{hermes2012g2}.

Our MA22 results for $\Gamma_2$ together with data from \text{E143}~\cite{Abe:1998wq}, \text{E155}~\cite{E155pdg2}, \text{HERMES2012}~\cite{hermes2012g2}, \text{RSS}~\cite{Slifer:2008xu}, \text{E01012}~\cite{Solvignon:2013yun} groups
for proton, deuteron and neutron
are listed in table~\ref{tab:BC}. The low-$x$ behaviour of $g_2$ which is not yet precisely measured, has considerable effect on any conclusion which we might be get.}

{{} The BC sum rule can be obtained analytically from  the covariant parton model as it is discussed in \cite{cpm1-1}.}
%
%
\begin{table*}[!htb]
	\caption{ \label{tab:BC} Comparison of the result of BC sum rule for $\Gamma_2^p$, $\Gamma_2^d$ and $\Gamma_2^n$ with world data from \text{E143}~\cite{Abe:1998wq}, \text{E155}~\cite{E155pdg2}, \text{HERMES2012}~\cite{hermes2012g2}, \text{RSS}~\cite{Slifer:2008xu}, \text{E01012}~\cite{Solvignon:2013yun}. }
	\begin{ruledtabular}
		\begin{tabular}{lccccccc}
			
			& \textbf{E143}~\cite{Abe:1998wq}  & \textbf{E155}~\cite{E155pdg2}  & \textbf{HERMES2012}~\cite{hermes2012g2}& \textbf{RSS}~\cite{Slifer:2008xu} & \textbf{E01012}~\cite{Solvignon:2013yun} & \textbf{KTA17}~\cite{Khanpour:2017cha}& \textbf{MA22}   \\
			& $0.03 \le x \le 1 $  & $0.02 \le x \le 0.8 $   & $0.023 \le x \le 0.9$  &$0.316 < x < 0.823 $&$0 \le x \le 1 $& $0.03 \le x \le 1 $& $0.03 \le x \le 1 $\\
			& $Q^2=5$ GeV$^2$& $Q^2=5$ GeV$^2$& $Q^2=5$ GeV$^2$& $Q^2=1.28$ GeV$^2$&$Q^2=3$ GeV$^2$&$Q^2=5$ GeV$^2$&$Q^2=5$ GeV$^2$   \\
			\hline  \hline
			$\Gamma_2^p$  & $-0.014 \pm 0.028$ &$-0.044 \pm 0.008$ & $0.006 \pm 0.029$ & $-0.0006 \pm 0.0022$&...&  $ -0.0196 \pm 0.0011$ &  $ -0.01554 \pm 0.00033$  \\
			$\Gamma_2^d$  & $-0.034 \pm 0.082$ &$-0.008 \pm 0.012$ &-  &$-0.0090 \pm 0.0026$&...& $ -0.0036 \pm 0.0005$ & $ -0.00401 \pm0.00006$   \\
			$\Gamma_2^n$ &-&-&-&$-0.0092 \pm 0.0035$&$0.00015 \pm 0.00113$& $ ~0.0060 \pm 0.0001$& $ ~0.00721 \pm 0.00033$  \\
		\end{tabular}
	\end{ruledtabular}
\end{table*}

\subsection{Efremov-Leader-Teryaev (ELT) Sum Rule}
{{} Considering the valence part of $g_1$ and $g_2$ structure functions and integrating them over $x$ variable the Efremov-Leader-Teryaev (ELT) sum rule is obtained. The ELT sum rule is derived like the Bjorken sum rule since the sea quarks are assumed to be identical in protons and neutrons. Hence it appears as:}
\begin{eqnarray}
	&&\int_0^1 dx ~ x[g_1^V(x) + 2 g_2^V (x)]=  \nonumber \\
	&&\int_0^1 dx ~ x[g_1^p(x) - g_1^n(x) + 2(g_2^p(x) - g_2^n(x))]=0.
\end{eqnarray}
{{} This sum rule is  only valid in the case of massless quarks and receives corrections from the quark mass but under presence of target mass corrections is preserved  \cite{Blumlein:1996vs}. {{} Like the BC sum rule, the ELT sum rule can be obtained by analytical considerations of CPM. More details can be found in \cite{cpm1-1}.}

By combining the data of E143~\cite{Abe:1998wq} and E155~\cite{E155pdg2} the numerical value for this sum rule  at Q$^2$=5 GeV$^2$ is $-0.011 \pm 0.008$  and what we obtain at the same energy scale  would be $0.01017\pm0.00004$.}
\section{Comparison for the spin structure functions}\label{helicity}
{{} Since our QCD analysis  has been validated by extracting the PPDFs via the fitting processes and also obtaining their evolved outputs and in continuation by considering several sum rules, we are now at the position to investigate the polarized structure functions. In this regard, we first back to what we got before. Our results,  MA22  PPDFs, as a function of $x$ at Q$_0^2=$ 1 GeV$^2$ along with the corresponding uncertainty bounds, is presented in Fig.~\ref{fig:partonqcdhttmc}.

The evolution of MA22 polarized parton distributions for a selection of $Q^2$ values indicates in Fig.\ref{fig:partonQ} while for comparison various parameterizations of KTA17~\cite{Khanpour:2017cha},KATAO11~\cite{Khorramian:2010qa},TKAA16~\cite{Shahri:2016uzl},NAAMY21~\cite{Nematollahi:2021ynm} at the NLO approximation are illustrated there. It is seen that by  increasing $Q^2$, except for the gluon density, the evolution of all  distributions tends to flatten out the peak.}

{{} Now for the  structure functions,  we see  that in different panels of Fig.\ref{fig:xg1p}, our MA22  predictions for the polarized structure functions of the proton $xg_1^p (x, Q^2)$, neutron $xg_1^n(x, Q^2)$ and deuteron $xg_1^d(x, Q^2)$  are compared  with respect to the fixed-target DIS experimental data from E143.}
 As we mentioned, MA22 refers to `pQCD+TMC+HT' scenario. The results from KATAO11 analysis in NLO approximation~\cite{Khorramian:2010qa}, TKAA16 analysis in NNLO approximation~\cite{Shahri:2016uzl},  KTA17 analysis in NNLO approximation ~\cite{Khanpour:2017cha}, THK14 analysis in NLO approximation~\cite{TaheriMonfared:2014var} and finally NAAMY21 analysis in NLO approximation~\cite{Nematollahi:2021ynm} are also depicted there. {{} We find our results are in good agreement with the experimental data  and in accord with other determinations over the entire range of $x$ at $Q^2$=5 GeV $^2$.

Further illustrations of the fit quality are presented in {{} different panels of Fig.\ref{fig:xg2p}}, for the $x g_2^{i = p, n, d}(x, Q^2)$ polarized structure functions, obtained from Eq.~\eqref{eq:g1full}.
In comparison with the $g_1$ data, the $g_2$ data have generally larger uncertainties which indicates the lack of knowledge for the $g_2$ structure function.
At the current level of accuracy,  MA22  is in agreement with data within their uncertainties. We need to a large number of data with higher precision to get a precise quantitative extraction of the $x g_2(x, Q^2)$. In fact we concentrate on the general characteristic of the $x g_2(x, Q^2)$ structure function.

Fig.~\ref{fig:xg1pjlab} is presenting our MA22 prediction for the polarized structure functions of the proton $xg_1^p (x, Q^2)$ while a comparison with the fixed-target DIS experimental data from JLAB17~\cite{Fersch:2017qrq} is  done there.

Fig.~\ref{fig:xg1t3Q1} represents our $xg_1^{\tau 3}(x,Q^2)$ with the results from  LSS ~\cite{Leader:2006xc} and JAM~\cite{Jimenez-Delgado:2013boa} groups. Analysis of the LSS group is based on  splitting the measured $x$ region into seven bins to determine the HT correction to $g_1$. The HT contribution  has been extracted by LSS group  in a model-independent way while its scale dependence is ignored. On the other side
an analytical form for the twist-3 part of $g_2$ is parameterized by the JAM group  where using integral relation of Eq.(\ref{eq:g1HT}) they calculated $g_1^{\tau 3}$ at the NLO accuracy in a global fit.

 E143 collaboration at SLAC reported the twist-3 contribution to proton spin structure function $x g_2^p$ structure function with relatively large errors ~\cite{Abe:1998wq}. We employ them and present our MA22 results for twist-3 part of $g_2$ in Fig.~\ref{fig:xg2twQ1} which are accompanied with  those of JAM~\cite{Jimenez-Delgado:2013boa} and BLMP~\cite{Braun:2011aw} groups.

However, within experimental precision the $g_2$ data are well described by the twist-2 contribution  but the precision of the current data is not sufficient enough to distinguish model precision. Hence we compute twist-3 part of $g_2$ for different targets  and depict them in Fig.\ref{fig:xg2tw3} which has significant contribution even at large Q$^2$ values.

In continuation to have a comparison, we compute the $xg_{\rm 1}^{\tau_3}$ and  indicate them in Fig.\ref{fig:xg1tw3}. We find that these functions vanish rapidly at $Q^2 > 5$ GeV$^2$  where in the limit of Q$^2 \rightarrow,  \infty$,  the $xg_{\rm 2}^{\tau_3}$ remains nonzero.

Up here we focused on longitudinal  polarized parton densities and structure functions. In next section we utilize  our MA22 analysis which we have done before to illustrate the transversal case which are including the polarized TMDs.}

\begin{figure}[!htb]
	\begin{center}
		\includegraphics[clip,width=0.5\textwidth]{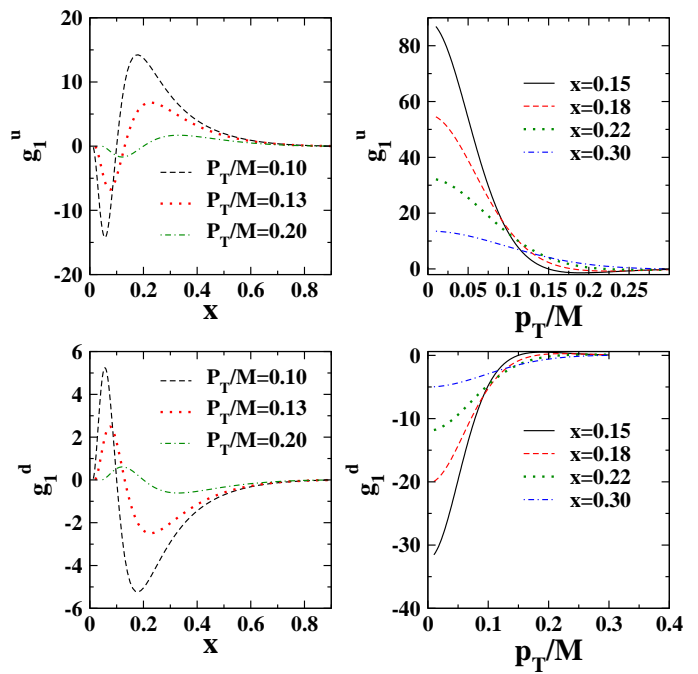}
		\caption{{\small   The TMD $g_{1}^{q}(x,\mathbf{p_{T}})$ for $u$-
				(\textit{upper panel}) and $d$-quarks (\textit{lower panel}). \textbf{Left
					panel}: $g_{1}^{q}(x,\mathbf{p_{T}})$ as function of $x$ for $p_{T}/M=0.10$
				(dashed), 0.13 (dotted), 0.20 (dash-dotted line). \textbf{Right panel}: $g_{1}%
				^{q}(x,\mathbf{p_{T}})$ as function of $p_{T}/M$ for $x=0.15$ (solid), 0.18
				(dashed), 0.22 (dotted), 0.30 (dash-dotted line).\label{ff3}}}
	\end{center}
\end{figure}
\begin{figure}[!htb]
	\begin{center}
		\includegraphics[clip,width=0.5\textwidth]{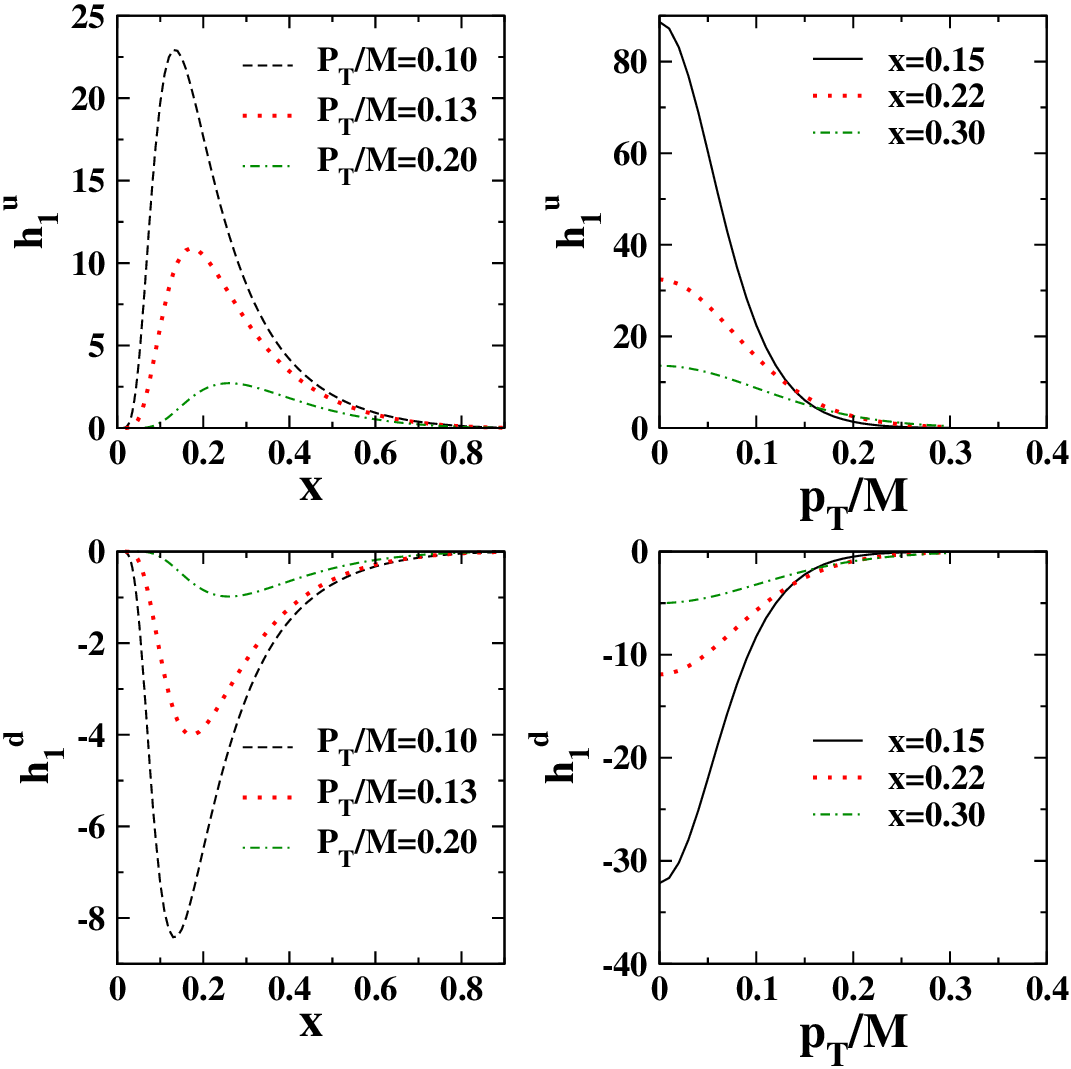}
		\caption{{\small  $h_{1}^{q}(x,\mathbf{p_{T}})$, for $u$- and $d$-quarks. \textbf{\ Left
					panel}: The TMDs as functions of $x$ for $p_{T}/M=0.10$ (dashed), 0.13
				(dotted), 0.20(dash-dotted lines). \textbf{Right panel}: The TMDs as functions
				of $p_{T}/M$ for $x=0.15$ (solid), 0.22 (dotted), 0.30
				(dash-dotted lines).%
				\label{ff4}}}
	\end{center}
\end{figure}
\begin{figure}[!htb]
	\begin{center}
		\includegraphics[clip,width=0.5\textwidth]{gudT.eps}
		\caption{{\small   $g_{1T}^{\bot q}(x,\mathbf{p_{T}})$,
				for $u$- and $d$-quarks. \textbf{\ Left
					panel}: The TMDs as functions of $x$ for $p_{T}/M=0.10$ (dashed), 0.13
				(dotted), 0.20(dash-dotted lines). \textbf{Right panel}: The TMDs as functions
				of $p_{T}/M$ for $x=0.15$ (solid), 0.22 (dotted), 0.30
				(dash-dotted lines).%
				\label{ff2}}}
	\end{center}
\end{figure}

\begin{figure}[!htb]
	\begin{center}
		\includegraphics[clip,width=0.5\textwidth]{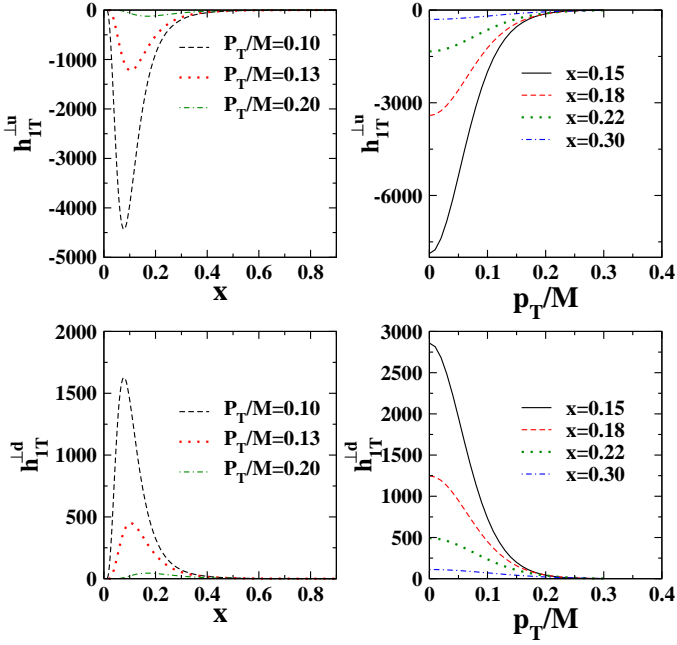}
		\caption{{\small
				$h_{1T}^{\perp q}(x,\mathbf{p_{T}})$ for $u$- and $d$-quarks. \textbf{\ Left
					panel}: The TMDs as functions of $x$ for $p_{T}/M=0.10$ (dashed), 0.13
				(dotted), 0.20(dash-dotted lines). \textbf{Right panel}: The TMDs as functions
				of $p_{T}/M$ for $x=0.15$ (solid), 0.18 (dashed), 0.22 (dotted), 0.30
				(dash-dotted lines).%
				\label{ff1}}}
	\end{center}
\end{figure}
\section{Predictions for polarized TMDs}\label{TMD}
{{} Since we achieved to sufficient information on longitudinal polarized  parton distributions and structure function, we are now at a situation to utilize the covariant parton model \cite{cpm1,cpm2} and extract the transverse momentum dependent (TMD) distributions in polarized case. Indeed  TMDs provide us new insight toward  a more complete understanding of the quark-gluon structure in a nucleon \cite{Collins:2003fm,Collins:2007ph,Collins:1999dz,Collins:2000gd,Hautmann:2007uw,Mulders:1995dh,Bacchetta:2006tn}. Without a more accurate and
realistic picture in  three dimensions of the nucleon which  includes  naturally transverse
motion, it would be hard to explain some experimental observations. In fact TMDs provide such pictures  and their necessities feel  more and more in nucleon investigations.

The first and simplest example of quark TMD is $f_1^q( x , k_ T )$. It arises when an unpolarized beam scatters off an unpolarized target hadron, and therefore does not carry quark/hadron spin information. The function $f_{1}^{q}(x,k_{T})$ provides the probability that a beam particle strikes a target quark of momentum fraction $x$ and transverse momentum $k_T$. It is related to the traditional DIS PDF $f_{1}^{q}(x)$  by $\int d^{2}k_{T}~f_{1}^{q}(x,k_{T})=f_{1}^{q}(x)$.

Similarly to $f_{1}^{q}(x,k_{T})$, we get the $g_{1}^{q}(x,k_{T})$ as longitudinal polarized TMD  and $ h_{1}^{q}(x,k_{T})$ as transverse polarized TMD, whose integrals are denoted respectively by $g_{1}^{q}(x)$ (presented before by $\Delta {q_i}(x)$) and $h_{1}^{q}(x)$ that we know them as quark longitudinal polarized (helicity) distribution  and the quark transversity distribution.

In addition to the three above TMDs  for quarks which are direct extension of the DIS PDFs, there are five other quark TMDs which depend not only on the magnitude of $k_T$ , but also on its direction. Therefore these TMDs vanish if simply integrated over $k_{T}$, and do not directly connect to DIS PDFs. They are:

1- The Sivers distribution $f_{1T}^{\bot ,q}$ which expresses, in a transversely polarized hadron, the asymmetric distribution of the quark transverse momentum, $p_z$, around the center of the $p_{x}$ and $p_{y}$ plane \cite{sivers}. The appearance of azimuthal asymmetric quark distribution in the transverse momentum space is often called the ``Sivers effect''. This TMD  has opposite signs in  semi-inclusive DIS (SIDIS) with respect to Drell-Yan processes  and it is therefore an odd time reversal function(T-odd function).

2- The Boer-Mulders function $h_{1}^{\bot ,q}(x,k_{T})$  characterizes the distribution of longitudinal polarized quarks in an unpolarized hadron \cite{B-M1}. It is also a T-odd function, like $f_{1T}^{\bot ,q}$. The rest tree TMDs are:

3-Function $h_{1T}^{\bot ,q}(x,k_{T})$  which is describing a transverse polarized quark inside a   transverse polarized  nucleon while its direction is perpendicular to a polarized nucleon. It is called  Pretzelosity function.

4-Function $g_{1T}^{\bot ,q}(x,k_{T})$  that is describing  the longitudinal  polarized quark inside a  transverse polarized nucleon and is named as  Worm-gear-I  function.
And finally:

5- Worm-gear-II function, denoted by $h_{1L}^{\bot ,q}(x,k_{T})$  and is describing the transverse  polarized quark inside a  longitudinal polarized nucleon,

Similarly to quark TMDs, gluon TMDs allow access to the gluonic orbital angular momentum, another possibly important contribution to the nucleon spin. Just as there are eight TMDs for quarks, there are eight gluon TMDs \cite{me}. Gluon TMDs were first proposed in 2001 \cite{ge}.

Here we only consider the Quark TMDs  that are twist-2 naively and time-reversal even
(T-even) functions. They have been  extracted via covariant parton model (CPM)  which is based on the Lorentz invariance and the assumption of a rotationally symmetric
distribution of parton momenta in the nucleon rest frame \cite{Efremov:2009ze}.

 {{} As a result of CPM, T-even polarized TMDs}
can be obtained at the leading twist approximation, in terms of a single \textquotedblleft generating
function\textquotedblright\ $K^{q}(x,\mathbf{p}_{T})$. They are given by~\cite{Efremov:2011ye,Bastami:2020rxn}
\begin{equation}
	\renewcommand{\arraystretch}{2.2}%
	\begin{array}
		[c]{rcrcl}%
		g_{1}^{q}(x,\mathbf{p}_{T}) & = & \displaystyle\frac{1}{2x}\left(  \left(
		x+\frac{m}{M}\right)  ^{2}-\frac{\mathbf{p}_{T}^{2}}{M^{2}}\right)  & \times &
		K^{q}(x,\mathbf{p}_{T})\;,\\
		h_{1}^{q}(x,\mathbf{p}_{T}) & = & \displaystyle\frac{1}{2x}\left(  x+\frac
		{m}{M}\right)  ^{2} & \times & K^{q}(x,\mathbf{p}_{T})\;,\\
		g_{1T}^{\perp, q}(x,\mathbf{p}_{T}) & = & \displaystyle\frac{1}{x}\left(
		x+\frac{m}{M}\right)  \; & \times & K^{q}(x,\mathbf{p}_{T})\;,\\
		h_{1L}^{\perp, q}(x,\mathbf{p}_{T}) & = & \displaystyle-\,\frac{1}{x}\left(
		x+\frac{m}{M}\right)  \; & \times & K^{q}(x,\mathbf{p}_{T})\;,\\
		h_{1T}^{\perp, q}(x,\mathbf{p}_{T}) & = & \displaystyle-\frac{1}{x}\, & \times
		& K^{q}(x,\mathbf{p}_{T})\;.
	\end{array}
	\label{Eq:all-TMDs}%
\end{equation}\\
According to \cite{Efremov:2009ze}) $K^{q}(x,\mathbf{p}_{T})$ as generating function is defined in compact notation by}
\begin{eqnarray}
	K^{q}(x,\mathbf{p}_{T})&=&M^{2}x\int\mathrm{d}\{p^{1}\} \\
	\mathrm{d}\{p^{1}\}&\equiv&\frac{\mathrm{d}p^{1}}{p^{0}}\;\frac{H^{q}(p^{0})}{p^{0}%
		+m}\;\delta\left(  \frac{p^{0}+p^{1}}{M}-x\right)  \,.
	\label{Eq:generating-function}%
\end{eqnarray}
{{} It can be shown} that due to rotational symmetry the following relations hold ~\cite{Efremov:2011ye}:

\begin{equation}
	K^{q}(x,\mathbf{p}_{T})=M^{2}\frac{H^{q}(\bar{p}^{0})}{\bar{p}^{0}+m}%
	,\qquad\bar{p}^{0}=\frac{1}{2}\,xM\,\left(  1+\frac{\mathbf{p}_{T}^{2}+m^{2}%
	}{x^{2}M^{2}}\right)  , \label{Eq:generating-function-2}%
\end{equation}
\begin{equation}
	\pi x^{2}M^{3}H^{q}\!\left(  \frac{M}{2}x\right)  =2\int_{x}^{1}%
	\frac{\mathrm{d}y}{y}\;g_{1}^{q}(y)+3\,g_{1}^{q}(x)-x\;\frac{\mathrm{d}%
		g_{1}^{q}(x)}{\mathrm{d}x}, \label{Eq:relation-g1-Hp}%
\end{equation}
{{}In deriving Eq.(\ref{Eq:relation-g1-Hp}) the limit $m\rightarrow0$ has been taken.
Consequently  the following result in that limit would be obtained  for the generating function ~\cite{Efremov:2011ye}:
\begin{widetext}
\begin{equation}
	K^{q}(x,\mathbf{p}_{T})=\frac{H^{q}(\frac{M}{2}\xi)}{\,\frac{M}{2}\xi}%
	=\frac{2}{\pi\xi^{3}M^{4}}\left(  2\int_{\xi}^{1}\frac{\mathrm{d}y}{y}%
	\;g_{1}^{q}(y)+3\,g_{1}^{q}(\xi)-x\;\frac{\mathrm{d}g_{1}^{q}(\xi)}%
	{\mathrm{d}\xi}\right)  ,\quad\xi=\,x\,\left(  1+\frac{\mathbf{p}_{T}^{2}%
	}{x^{2}M^{2}}\right)  . \label{e3}%
\end{equation}
\end{widetext}
Substituting the above relations in Eq.(\ref{Eq:all-TMDs}), the following result for the $g_{1}^{q}(x,\mathbf{p}_{T})$ would be obtained:
\begin{equation}
	g_{1}^{q}(x,\mathbf{p}_{T})=\frac{2x-\xi}{\pi\xi^{3}M^{3}}\left(  2\int_{\xi
	}^{1}\frac{\mathrm{d}y}{y}\;g_{1}^{q}(y)+3\,g_{1}^{q}(\xi)-\xi\;\frac
	{\mathrm{d}g_{1}^{q}(\xi)}{\mathrm{d}\xi}\right)  . \label{e4}%
\end{equation}
Based on above relation and using the MA22 analysis which we did in this paper for $g_{1}^{q}(x)$ at $4\,\mathrm{GeV}^{2}$ in the NNLO approximation, we could obtain the result for  $g_{1}^{q}(x,\mathbf{p}_{T})$ which has been shown in Fig.[\ref{ff3}] for $u$ and $d$ quarks.

Using Eq.(\ref{Eq:all-TMDs}) and in the limit $m\to0$ the other TMDs can be obtained. They are presented in below which which are different by simple $x$-dependent prefactors ~\cite{Efremov:2011ye}:
\begin{eqnarray}
	h_{1}^{q}(x,\mathbf{p}_{T}) =\frac{x}{2}K^{q}(x,\mathbf{p}_{T}),\nonumber\\ \;\;
	g_{1T}^{\perp, q}(x,\mathbf{p}_{T}) = K^{q}(x,\mathbf{p}_{T}),\;\;\nonumber\\
	h_{1T}^{\perp, q}(x,\mathbf{p}_{T}) =-\frac{1}{x}K^{q}(x,\mathbf{p}%
	_{T}).\label{e5}%
\end{eqnarray}
The result for $h_{1}^{q}(x,\mathbf{p}_{T})$  is depicted in Fig.~\ref{ff4}.  In  Fig.~\ref{ff2} the result for  $g_{1T}^{\perp q}$ with respect to $x$   and $p_T/M$  is shown. It does not need to plot $h_{1L}^{\perp q}$ since in the used approach  this TMD is equal to $-g_{1T}^{\perp q}$ \cite{Efremov:2009ze}.
As can be seen from Fig.\ref{ff3}, $g_{1}^{q}(x,\mathbf{p}_{T})$ is
the only TMD which has positive and negative values. The other TMDs in other figures do not change sign  which follows from Eqs.(\ref{Eq:all-TMDs},~\ref{e5}).

We should note that among all TMDs, as we see from  Fig.\ref{ff1}, $h_{1T}^{\perp q}(x,\mathbf{p}_{T})$ as pretzelosity function has largest absolute value which is due to the prefactor $1/x$. This function has its own worth since in some quark models
\cite{She:2009jq,Avakian:2010br}, including the utilized approach in \cite{Efremov:2010cy,Avakian:2010nz}, this function is related to
quark orbital angular momentum.}
\section{Conclusions} \label{Summary}
{{}
Determining the nucleon spin structure functions $g_1(x,Q^2)$ and $g_2(x,Q^2)$ and their moments is  the main goal of our present MA22  analysis.
They are essential to test QCD sum rules and to evaluate the TMDs. We provided a unified and consistent PPDF through an achievement, containing an excellent description of the fitted data while we employed TMC and HT effects in our analysis. Within the known very large uncertainties arising from the lack of constraining data, our helicity distributions are in good consistency with other extractions. Here the TMCs and HT effects, which are relevant in the region of low $Q^2$, have also been studied for the several sum rules at the NNLO approximation. Our results for the reduced matrix element $d_2$ at the NNLO approximation have also been presented.  We also studied Burkhardt-Cottinghan and  Efremov-Leader-Teryaev sum rules. To scrutiny them more accurate data are needed.

Finally we studied the behavior of the TMD structure functions which are time-reversal even with respect to $x$ and $p_T/M$  variables at the  NNLO approximation, based on the covariant parton model. { Our MA22 results, containing  analysis of up to date and last data on nucleon spin structure functions, with respect  to what we did in \cite{Khanpour:2017cha},} can be compared  with the results from \cite{Efremov:2011ye} which indicated adequate and acceptable behaviours.

This study can be extended to include other TMDS while higher twist effect is employed. We hope to report on this issue as our further research task. }
\newline

\section*{Acknowledgments}
The Authors are indebted P.~Zavda for reading the manuscript and providing useful comments. We are grateful to O.~V.~Treyaev  for his useful comments and suggestion.  We are appreciated P.~Schweitzer to read our manuscript and give us his opinion about it.  S.~A.~T. is thankful from  the School of Particles and
Accelerators, Institute for Research in Fundamental
Sciences (IPM) to make the required facilities to do this
project.  A.~M  acknowledges the Yazd university for the provided facility to do this project.	

%

\end{document}